\documentclass[prx, preprint, a4paper, amsmath, amssymb, amsfonts]{revtex4-2}

\usepackage{graphicx}
\usepackage{hyperref}
\usepackage{enumitem}

\newcommand{\nqp}{\ensuremath{n_\text{qp}}}
\newcommand{\qr}{\ensuremath{q_r}}
\newcommand{\fres}{\ensuremath{f_0}} % Resonant frequency
\newcommand{\fpmp}{\ensuremath{f_\text{p}}} % Pump frequency
\newcommand{\fsgnl}{\ensuremath{f_\text{s}}} % Signal frequency
\newcommand{\bw}{\ensuremath{{\Delta f}_r}} % 3dB bandwidth
\newcommand{\bwzero}{\ensuremath{{\Delta f}_0}} % 3dB bandwidth (r=0)
\newcommand{\xvec}{\ensuremath{\hat{\mathbf{x}}}}
\newcommand{\op}{\ensuremath{0}}
\newcommand{\kps}{\ensuremath{k_{p,s}}}
\newcommand{\ks}{\ensuremath{k_{s}}}
\newcommand{\as}{\ensuremath{a_{s}}}
\newcommand{\Cavendish}{Cavendish Laboratory, JJ Thomson Avenue, Cambridge, CB3 0HE, UK}

\newlength\mylen
\newlist{mycases}{enumerate}{1}
\setlist[mycases, 1]{label=\textbf{Case~\arabic*:}, labelwidth=\dimexpr-\mylen-\labelsep\relax ,leftmargin=0pt, align=right}

\begin{document}

\title{Effects of reactive, dissipative and rate-limited nonlinearity on the behaviour of superconducting resonator parametric amplifiers}
\date{\today}

\author{Christopher N. \surname{Thomas}}
\email[]{cnt22@cam.ac.uk}
\affiliation{\Cavendish}
\author{Stafford \surname{Withington}}
\affiliation{\Cavendish}
\author{Songyuan \surname{Zhao}}
\affiliation{\Cavendish}

\begin{abstract}
We present a formalism for modelling parametric amplification by resonators subject to rate-limited nonlinearity of mixed reactive/dissipative character, with relevance to superconducting devices.
The nonlinearity is assumed to be characterised by a single state parameter that responds to changes in the energy stored in the resonator with finite response time.
We show how the operating point and small signal amplification behaviour of the pumped resonator can be calculated, characterised and optimised in terms of a set of three dimensionless parameters.
The formalism is illustrated with a simple, first-order, model nonlinearity and the implications for amplification via quasiparticle generation in a superconductor discussed.
Throughout we describe how the parameters needed to characterise the device can be determined experimentally from steady-state measurements.
A key result of this paper is that rate-limiting of a nonlinear mechanism does not preclude amplification, although it does limit the bandwidth over which it may be achieved.
\end{abstract}


\maketitle

\section{Introduction}\label{sec:introduction}

There is demand for microwave amplifiers with near quantum-limited noise performance for readout of quantum devices such as qubits, as well as for experiments in fundamental physics such as axion haloscopes and measurements of the neutrino mass by cyclotron radiation emission spectroscopy.
Where only moderate fractional bandwidth is needed, parametric amplifiers made from nonlinear superconducting resonators are a competitive technology.
As well as delivering the necessary sensitivity, they are easily integrated with other superconducting components to realise more complicated systems, and can be operated in a phase-sensitive mode, allowing generation and amplification of squeezed states.
Although narrower band than their superconducting travelling counterparts\,\cite{eom2012wideband}, resonator amplifiers are typically easier to fabricate, have flatter passbands and require lower pump power.
They are also typically much smaller, allowing denser device integration on-chip.

Several sources of nonlinearity can be used to realise a superconducting resonator amplifier.
A popular approach has been to use composite superconducting devices such as SQUIDs as a source of nonlinear inductance\,\cite{aumentado2020superconducting} and amplifiers of the type, usually referred to as Josephson-Parametric Amplifiers (JPAs), are widely used for the readout of qubits and quantum memory devices\,\cite{mutus2013design,o2021random}.
However, there is increasing interest exploiting nonlinearities intrinsic to the superconducting films, so as to simplify fabrication and potentially improve power handling.
Efforts in this area have focused on nonlinear kinetic inductance\,\cite{tholen2007nonlinearities}, but other mechanisms, such as weak-links introduced either by granular microstructure\,\cite{abdo2006nonlinear} or microscopic patterning\,\cite{tholen2009parametric} are also of interest.
Speculatively, there is also the possibility of engineering nonlinearity by introducing nonlinear dielectrics, either as coating layers or as the substrate on which the resonator is fabricated.

Whether a source of nonlinearity can be used for parametric amplification under pumping and, if so, how to optimize for gain, bandwidth and power handling depends critically on three characteristics:
(i) The ratio of reactive to dissipative response.
(ii) The `large' signal behaviour of the nonlinearity and the corresponding accessible operating points under pumping.
(iii) What we will call the `speed' of the nonlinearity, corresponding to the maximum rate at which it can modulate device parameters.
Typical analyses of resonator amplifiers focus only on the ideal case: a purely-reactive, instantaneous, nonlinearity with simple functional form, e.g. quadratic dependence on a state variable.
This is instructive in building a qualitative understanding of amplifier operation, however it is straightforward to see how even small deviations from this ideal may change device behaviour significantly.
For example, any one of the characteristics on its own may prevent gain being achieved:
(i) The presence of nonlinear dissipation may alter the dynamics of the resonator in a such a way that gain becomes impossible.
(ii) An operating point with the necessary small signal parameters for gain may not exist.
(iii) The response time of the nonlinearity may be too slow to produce modulation of a device parameter at the rate necessary to couple signal and idler.
A model that accounts for all three characteristics is therefore crucial to quantitatively interpret data and for design optimization.

As an example, consider a device that uses nonlinearity in the surface impedance of a superconductor to achieve parametric amplification.
The electrodynamics of a superconductor are governed by the populations of Cooper pairs and quasiparticles\,\cite{annett2004superconductivity}.
Nonlinear behaviour of the surface impedance is usually attributed to the `intrinsic' nonlinearity in kinetic inductance\,\cite{pippard1950field,parmenter1962nonlinear,swenson2013operation,eom2012wideband}.
At a microscopic level this arises from modulation of the underlying quantum states\,\cite{semenov2016coherent} and leads to a fast nonlinearity that is predominately reactive at temperatures well below the superconducting critical temperature.
However, a superconductor in a microwave field is also subject to quasiparticle heating, whereby sub-gap photons are able to break Cooper pairs by an indirect process\,\cite{goldie2012non,de2014evidence}.
The loss of Cooper pairs affects the kinetic inductance and the generation of quasiparticles from pair-breaking leads to increased resistance, yielding a mixed reactive-dissipative nonlinearity in the surface impedance in the same temperature limit.
This nonlinearity is slow as it is limited by the quasiparticle relaxation time, which can be of the order of milliseconds in elemental superconductors such as Al\,\cite{mazin2020superconducting}.
Both mechanisms can in principle generate gain and we may distinguish which is at work, and so optimize performance appropriately, based on the differences in their reactive/dissipative ratios and speeds.

In this paper we present a framework for modelling parametric amplification in nonlinear resonators where the nonlinearity is of mixed reactive/dissipative type and rate-limited.
We consider the mode of operation where the pump, signal and idler all lie within the bandwidth of the same resonant mode.
The starting point is a large-signal model of the nonlinearity, derived either from theory or from measurements of the response of the resonator to a large signal tone, as discussed in our companion paper\,\cite{thomas2020nonlinear}.
In the first part of a paper we use a perturbation analysis to derive the corresponding behaviour as an amplifier when pumped.
In order to account for the speed of the nonlinearity (for what we believe is the first time) we introduce an intermediate, rate-limited, process via which modulation of the device parameters occurs.
The inclusion of this additional process necessitates a re-derivation of some standard results, and we take the opportunity to do so in a notation familiar to the superconducting resonator community.
In the second part of the paper, we present a number of illustrative simulations and discuss their relevance to superconducting devices.
We show how the `Duffing' type model of a nonlinear resonator\,\cite{jose1998classical,swenson2013operation} can be extended to model a mixed reactive/dissipative nonlinearity by introducing a phase-angle to parametrise the relative response, as well as demonstrating how a finite response time sets a limit on gain bandwidth product.
Our analysis is presented in a generic way, so as to allow its application to different nonlinearities and different resonator realisations.

\section{Device model}\label{sec:model}

\subsection{State-space representation of a nonlinear resonator coupled to a readout circuit}\label{sec:resonator_model}

\begin{figure}
\centering
\includegraphics[width=8cm]{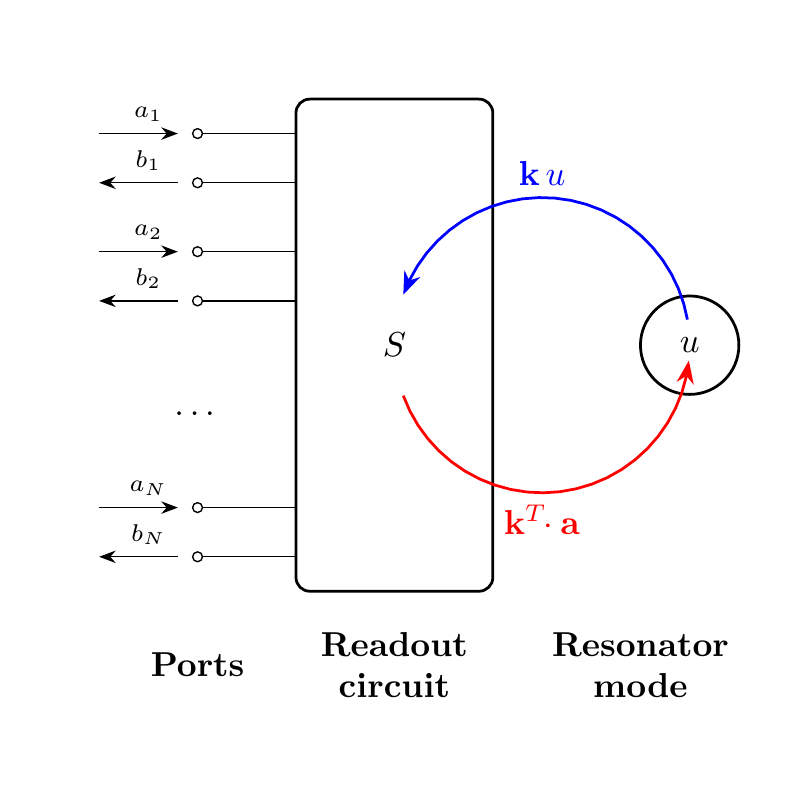}
\caption{\label{fig:resonator_model} Resonator model.}
\end{figure}

Our model comprises a single resonator coupled to an $N$-port microwave readout circuit, as shown in Figure \ref{fig:resonator_model}.
Because the amplification process is necessarily nonlinear, it is most easily analysed using a state-space approach.

For a state-space representation, it is necessary to identify input, output and internal dynamical variables.
To keep the analysis generic, we use variables normalized in terms of power and energy.
At each port $m$ of the coupling circuit we define input and output analytic signals $a_m(t)$ and $b_m(t)$, normalised so that $P_m (t) = |a_m(t)|^2 - |b_m(t)|^2$ is the time-averaged power flow into port $m$ at time $t$.
In the case of a transmission line port, $a(t)$ and $b(t)$ are the complex analytic-signal representations of the amplitudes of the incoming and outgoing power waves, respectively.
For brevity, we group the input signals into a $N$-element column vector $\mathbf{a}(t)$,  and likewise the output signals into an $N$-element column vector $\mathbf{b}(t)$.
To characterise the state of the internal resonator, we use the complex amplitude $u(t)$ of the resonant mode, normalised so that $U(t) = |u(t)|^2$ is the time-averaged total stored energy.

First, consider the case where the resonator and coupling circuit are fully decoupled.
Let $\fres$ and $Q_i$ be the resonant frequency and quality factor of the resonator in this limit.
We refer to $Q_i$ as the \textit{internal Q-factor}, as by definition it is associated solely with losses internal to the resonator.
We assume that in this same limit, the readout circuit is reciprocal, lossless and that its behaviour for quasi-monochromatic signals having frequencies near $\fres$ can be characterised by a \emph{constant} scattering matrix $\mathsf{S}_0$ defined by
\begin{equation}\label{eqn:def_S}
	\mathbf{b}(t) = \mathsf{S}_0 \cdot \mathbf{a}(t).
\end{equation}
The reciprocal and lossless conditions require $\mathsf{S}_0$ be symmetric ($\mathsf{S}_0^T = \mathsf{S}_0$, where $^T$ indicates matrix transposition) and unitary ($\mathsf{S}_0^\dagger = \mathsf{S}_0^{-1}$, where superscript $\dagger$ indicates conjugate transposition).

With the decoupled behaviour specified, for a linear resonator the state-space representation of the fully-coupled system follows immediately from temporal coupled mode theory\,\cite{zhao2019connection,haus1984waves}.
With the definitions above, the output and state equation are, respectively,
\begin{equation}\label{eqn:coupled_output_equation}
	\mathbf{b}(t) = \mathsf{S}_0 \cdot \mathbf{a}(t) + \mathbf{k} \, u(t),
\end{equation}
and
\begin{equation}\label{eqn:coupled_state_equation}
	\frac{du}{dt} = 2 \pi i \fres \biggl\{ 1+ \frac{i}{2 Q_i} + \frac{i}{2 Q_c} \biggr\} u(t)
	+ \mathbf{k}^T \! \cdot \mathbf{a} (t),
\end{equation}
where $\mathbf{k}$ is an $N$-element vector determined by the coupling and
\begin{equation}\label{eqn:def_qc}
	Q_c = 2 \pi \fres / |\mathbf{k}|^2.
\end{equation}
We will refer to $Q_\text{c}$ as the \emph{coupling quality factor} of the whole system, as it is associated with the power lost from the resonator through the now-coupled ports.
For (\ref{eqn:coupled_output_equation}) and (\ref{eqn:coupled_state_equation}) to hold, we must have
\begin{equation}\label{eqn:s_and_k_relationship}
	\mathsf{S}_0 \cdot \mathbf{k}^* = - \mathbf{k},
\end{equation}
as shown in Section \ref{sec:s_and_k_relationship} of the supplementary material, and discussed in detail in \cite{zhao2019connection}.
In addition, the frequency dependence of the coupling circuit must be weak over the bandwidth of the resonator.
This a reasonable assumption for the simple capacitive or inductive coupling circuits used in practice.
More general readout- and coupling-circuit behaviour can be accommodated by making $\mathsf{S}_0$ and $\mathbf{k}$ time-dependent linear operators, however it is not necessary to do so for our aim of studying the fundamentals of parametric amplification.
Section \ref{sec:hw_resonator} of the supplementary material provides a derivation of (\ref{eqn:coupled_output_equation}) and (\ref{eqn:coupled_state_equation}) for the concrete example of an end-coupled transmission line resonator.

A nonlinear resonator can be modelled by modifying the right-hand sides of (\ref{eqn:coupled_output_equation}) and (\ref{eqn:coupled_state_equation}).
A wide range of behaviour is possible, but most practical amplifiers satisfy two simple restrictions.
First, the coupling is unaffected by the nonlinearity, and so the coupling vector $\mathbf{k}$ remains constant and (\ref{eqn:coupled_output_equation}) is unchanged.
Second, the effect of the nonlinearity can be modelled as time-dependent shifts $\delta \fres (t)$ and $\delta Q^{-1}_i (t)$ in the resonant frequency and internal dissipation factor (reciprocal of the internal quality factor), with (\ref{eqn:coupled_state_equation}) becoming
\begin{equation}\label{eqn:nl_coupled_state_equation}
	\frac{du}{dt} = 2 \pi i \fres \biggl\{ 1+ \frac{i}{2 Q_r}
	+ \frac{\delta \fres (t)}{\fres}
	+ \frac{i \delta Q_i^{-1} (t)}{2} \biggr\} u(t)
	+ \mathbf{k}^T \! \cdot \mathbf{a} (t),
\end{equation}
where $Q_r$ is the total Q-factor in the zero power limit, as given by $Q_r^{-1} = Q_i^{-1} + Q_c^{-1}$.
Through (\ref{eqn:nl_coupled_state_equation}) we can define, in a rigorous way, a nonlinearity as having a reactive effect if it contributes to $\delta \fres (t)$, and as having a dissipative effect if it contributes to $\delta Q_i^{-1} (t)$.
Thus in our model, parametric changes in the underlying circuit are described by changes in the resonant frequency and Q-factor, rather than by changes in the inductance, capacitance, and resistance of some assumed equivalent circuit, providing a more general description.

\subsection{Rate-limited parametric behaviour}\label{sec:model_nonlinearity}

According to (\ref{eqn:nl_coupled_state_equation}), the complex pole describing the resonator can be perturbed slowly in the complex plane.
The next step is to specify the assumed forms of the shifts in resonant frequency and dissipation factor.
In particular, we will introduce the idea that the physical process that drives the dominant nonlinearity can only produce a rate-limited modulation of the resonator's parameters.
We will study a generalised, first-order model based on two assumptions: First, there is a real-valued state variable $v$, descriptive of the physical process, which drives the shift in $\delta \fres (t)$ and $\delta Q_i^{-1} (t)$ through a complex-valued function $g(v)$:
\begin{equation}\label{eqn:model_non_linearity}
	\frac{\delta \fres}{\fres}
	+ \frac{i \delta Q_i^{-1}}{2} = \frac{g(v)}{Q_r},
\end{equation}
where $Q_r$ is included as a normalising factor to simplify the algebra later.
Second, the dynamical behaviour of the new variable is related to the time-averaged total energy $U(t)$ stored in the resonator by a differential equation of the form:
\begin{equation}\label{eqn:v_state_equation}
	\frac{dv}{dt} = h(v, U).
\end{equation}
In this approach, $g(\nu)$ controls the nature of the nonlinearity, and $h(v, U)$ controls the dynamical behaviour of the solid-state process that drives the nonlinearity.
Specifically, the real part of $g(\nu)$ controls the reactive response and the imaginary part the dissipative part, allowing both behaviours simultaneously.
The function $h(v, U)$ has a dual role.
Obviously, it determines the dynamical behaviour of $v$ in response to changes in the energy in the resonator.
However, since $v$ must be constant in the steady state, its roots, as defined by solutions of $h(v, U) = 0$, also determine the operating point(s) of the device.
The rate limiting of the parametric modulation that occurs through the temporal filtering of (\ref{eqn:v_state_equation}) will be made explicit in the upcoming analysis.

(\ref{eqn:model_non_linearity}) and (\ref{eqn:v_state_equation}) can be used to model a wide-range of processes in superconducting resonators.
The most obvious are processes influenced directly by $U$, e.g. saturation of two-level-systems in the dielectric components of the device\,\cite{gao2008physics}.
However, they can also model processes that involve a flow of power from the resonator to another component of the system, the excitation of which in turn affects the resonator.
In this latter case the power flow can be expressed in the form $U/Q(v)$, where $Q(v)$ the associated quality factor contribution from the process.
Examples include physical heating of the device (where $v$ would be a temperature) and quasiparticle generation processes (where $v$ would represent a quasiparticle excess and the power flow is the energy needed to break Cooper pairs at the necessary rate).

In choosing the dependence to be based on the `time-average' total energy, $U(t) = |u(t)|^2$, we have limited ourselves to considering `slow' processes.
This is consistent with our assumption that the processes of interest are rate-limited.
`Fast' processes exist that depend on $u(t)^2$, i.e. the square of the complex amplitude of the field mode of the resonator.
In this case, a resonator's parameters can be modulated at frequencies significantly greater than the pump and signal frequencies themselves.
Although these effects can be included, we shall not, for brevity, discuss them here.

\subsection{Active component of the signal vector}\label{sec:ss_scalar_eqn}

Figure \ref{fig:resonator_model} shows $2N$ travelling waves scattering and interacting with a resonator.
Because the resonator has only one complex degree of freedom, only one input vector can interact with the resonator, and so only one input vector can be active, and the other must be passive.
In this section, we will derive an alternative form of (\ref{eqn:coupled_output_equation}) and (\ref{eqn:nl_coupled_state_equation}) where the components of the input and output signals that interact with the resonator are separated from those that do not.

Define a new complex unit vector $\xvec$ by the relation
\begin{equation}\label{eqn:def_x_vector}
	\mathbf{k} = i \kappa \xvec^*
\end{equation}
for $\kappa = |\mathbf{k}|$.
The reason for defining a separate variable to denote $|\mathbf{k}|$ will become apparent shortly.
(\ref{eqn:s_and_k_relationship}) implies
\begin{equation}\label{eqn:s_and_x_relationship}
	\mathsf{S}_0 \cdot \xvec = \xvec^*.
\end{equation}
(\ref{eqn:s_and_x_relationship}) constrains $\xvec$ but does not define it uniquely, so we must find $\xvec$ either by circuit analysis (via $\mathbf{k}$) or by measurement.
Exploiting the fact $\mathsf{S}_0$ is symmetric, then for an arbitrary vector $\mathbf{v}$,
\begin{equation}\label{eqn:xtSv}
	\xvec^T \! \cdot \mathsf{S}_0 \cdot \mathbf{v}
	= (\mathsf{S}_0 \cdot \xvec)^T \cdot \mathbf{v}
	= \xvec^\dagger \cdot \mathbf{v}.
\end{equation}
From which we derive two special cases: First, setting $\mathbf{v} = \mathbf{x}$, gives $\xvec^T \! \cdot \mathsf{S}_0 \cdot \mathbf{x} = 1$.
Second, if $\mathbf{v}$ is orthogonal to $\xvec$, so $\xvec^\dagger \cdot \mathbf{v} = 0$, then $\mathsf{S}_0 \cdot \mathbf{v}$ is orthogonal to $\mathbf{x}^*$.

The desired alternate form for (\ref{eqn:coupled_output_equation}) and (\ref{eqn:nl_coupled_state_equation}) follows by considering the decomposition of $\mathbf{a}$ into components parallel and perpendicular to $\xvec$.
Explicitly, let
\begin{equation}\label{eqn:a_decomp}
	\mathbf{a} (t) = \mathbf{a}' (t) + a(t) \xvec,
\end{equation}
where
\begin{equation}\label{eqn:def_a_s}
	a(t) = \xvec^\dagger \cdot \mathbf{a} (t)
\end{equation}
is the component parallel to $\xvec$ and $\mathbf{a}'$ is the perpendicular component, so that $\xvec^\dagger \cdot \mathbf{a}'(t) = 0$.
Similarly, define a new reflected signal $b(t)$ corresponding to the component of $\mathbf{b} (t)$ parallel to $\xvec^*$ by
\begin{equation}
	b(t) = \xvec^T \cdot \mathbf{b}(t).
\end{equation}
Consider substituting (\ref{eqn:a_decomp}) and (\ref{eqn:def_x_vector}) into the original output equation, (\ref{eqn:coupled_output_equation}).
By taking the scalar product of the result with $\xvec^T$ and making use of the special cases of (\ref{eqn:xtSv}), an output equation for $b(t)$ in terms of $u(t)$ and $a(t)$ can be obtained.
This result can, in turn, be subtracted from (\ref{eqn:coupled_output_equation}) to find the component of $\mathbf{b}$ perpendicular to $\xvec^*$.
Doing so and rewriting (\ref{eqn:nl_coupled_state_equation}) in terms of the new variables, we may re-express (\ref{eqn:coupled_output_equation}) and (\ref{eqn:nl_coupled_state_equation}) as
\begin{equation}\label{eqn:b_vec_from_b}
	\mathbf{b}(t) = \mathsf{S}_0 \cdot \mathbf{a}'(t) + b(t) \xvec^*,
\end{equation}
\begin{equation}\label{eqn:scalar_output_equation}
	b(t) = a(t) + i \kappa u(t)
\end{equation}
and
\begin{equation}\label{eqn:scalar_state_equation}
	\frac{du}{dt} = 2 \pi i \fres \biggl\{ 1+ \frac{i}{2 Q_r}
	+ \frac{\delta \fres (t)}{\fres}
	+ \frac{i \delta Q_i^{-1} (t)}{2} \biggr\} u(t)
	+ i \kappa a (t).
\end{equation}

From (\ref{eqn:b_vec_from_b})--(\ref{eqn:scalar_state_equation}), we see that only a single component, $a(t)$, of the input signal couples to the resonator, which in turn affects only a single component, $b(t)$, of the output signal.
The behaviour, described by (\ref{eqn:scalar_output_equation}) and (\ref{eqn:scalar_state_equation}), is that of a resonator coupled to a single port by a scalar coupling coefficient $\kappa$ (note $\kappa^2 = |\mathbf{k}|^2 = 2 \pi \fres / Q_c$ by definition).
The other components of the input signal, represented by $\mathbf{a}'(t)$, are unaffected by the resonator's presence and simply see the scattering matrix of uncoupled readout circuit.
The physical picture is an underlying, active, single-port device embedded in a multi-port circuit.
As such, the vectorial nature of the problem is imposed purely by the choice of coupling circuit.
The behaviour of the underlying single port device can be characterised in terms of scalar reflection coefficients.

Going forward, it will be more convenient to work in terms of the active degrees of freedom $a(t)$ and $b(t)$ only, as described by (\ref{eqn:scalar_output_equation}) and (\ref{eqn:scalar_state_equation}).
However, throughout the paper, we relate the results back to the measurable vector-valued variables $\mathbf{a} (t)$ and $\mathbf{b}(t)$.
To this end, it is useful to note that 
\begin{equation}
	\mathbf{S}_0 \cdot \mathbf{a}' (t)
	= \mathsf{S_0} \cdot ( \mathbf{a} (t) - a (t) \xvec )
	= \mathsf{S_0} \cdot \mathbf{a} (t) - a (t) \xvec^*
\end{equation}
so (\ref{eqn:b_vec_from_b}) can be written as
\begin{equation}\label{eqn:alt_output_equation}
	\mathbf{b} (t)
	= \mathsf{S}_0 \cdot \mathbf{a} (t)
	- \{ a(t) - b(t) \} \xvec^*.
\end{equation}
This form will prove particularly convenient when calculating the scattering parameters of the full circuit in the frequency domain.
(\ref{eqn:alt_output_equation}) also shows that the active degrees of freedom can be accessed experimentally via the difference between the measured output signal and that expected from the the readout circuit alone, $\mathbf{b} - \mathsf{S}_0 \cdot \mathbf{a}$, as might have been expected intuitively.
Typically $\mathsf{S_0}$ and $\mathbf{k}$ (and therefore $\xvec$) will be known a priori from circuit analysis.
If not $\mathsf{S}_0$, can usually be inferred from off-resonance measurements.
Once $\mathsf{S}_0$ is known, (\ref{eqn:alt_output_equation}) implies $\xvec$ can then be inferred from low-power measurements near resonance, provided the output signals at all the ports can be recorded.

\section{Operation as an amplifier}\label{sec:amplification}

The nonlinear resonator described in Section\,\ref{sec:model_nonlinearity} can potentially operate as a parametric amplifier if it is driven into a nonlinear regime with a sufficiently strong `pump' at frequency $\fpmp$.
If a much weaker input signal of frequency of frequency $\fsgnl$ is applied simultaneously, nonlinear effects can produce a linearly amplified output at the `signal' frequency $\fsgnl$, along with a new component at the `idler' frequency $2 \fpmp - \fsgnl$.

In this section we analyse this mode of operation.
First we consider the response of the resonator to the pump alone, which determines the resonator's operating point.
We then carry out a small-signal analysis about this operating point to derive expressions for the signal and idler gains.
Finally, we consider how the quantities needed to predict these gains can be obtained from swept-frequency/power measurements of the steady state scattering parameters of the resonator.

\subsection{Response to the pump}\label{sec:response_to_pump}

Assume a pump of the form
\begin{equation}\label{eqn:pump_signal}
	\mathbf{a}(t) = \mathbf{a}_0 e^{2 \pi i f_p t},
\end{equation}
is applied to the resonator, where $\mathbf{a}_0$ is a constant.
Given sufficient time the resonator will settle into a steady state where the nonlinear state parameter and total stored energy are constants: $v (t) = v_0$ and $|u(t)|^2 = U_0$.
Following (\ref{eqn:v_state_equation}), $v_0$ and $U_0$ must satisfy the operating point equation
\begin{equation}\label{eqn:op_point_equation}
	h(v_0, U_0) = 0.
\end{equation}
Similarly, we would expect $u(t)$ and $\mathbf{b}(t)$ to have the form
$\mathbf{b}(t) = \mathbf{b}_0 \exp(2 \pi i f_p t)$ and $u(t) = u_0 (t) = u_0 \exp(2 \pi i f_p t)$,
where $\mathbf{b}_0$ and $u_0$ are constants and $|u_0|^2 = U_0$.
Let $a_0 =  \xvec^\dagger \cdot \mathbf{a}_0$ and $b_0 =  \xvec^T \cdot \mathbf{b}_0$, then we can define an effective complex reflection coefficient for the active field components in the steady state by $\Gamma_p = b_0 / a_0$.
Equivalently, we can define a scattering matrix $\mathsf{S}_p$ for the whole system in the steady state by $\mathbf{b}_0 = \mathsf{S}_p \cdot \mathbf{a}_0$.

Expressions for $\mathbf{b}$ and $u_0$ in terms of $\mathbf{a}_0$ can be found by solving (\ref{eqn:b_vec_from_b})--(\ref{eqn:scalar_state_equation}) for the assumed signal forms.
We then find
\begin{equation}\label{eqn:u0}
	u_0 = \frac{Q_r}{\pi f_0} \frac{i \kappa a_0}{1 + 2 i \{ y_p - g(v_0) \}},
\end{equation}
\begin{equation}\label{eqn:gamma_p}
	\Gamma_p = 1 - \frac{2Q_r}{Q_c} \frac{1}{1 + 2 i \{ y_p - g(v_0) \}}
\end{equation}
and
\begin{equation}\label{eqn:sp}
	\mathsf{S}_p = \mathsf{S}_0 - (1 - \Gamma_p) \, \xvec^* \xvec^\dagger.
\end{equation}
Here $y_p = Q_r ( \fpmp - \fres) / \fres$ is the detuning of the pump frequency from the unpumped resonant frequency, measured in linewidths (in the limit of zero applied power).
Expressing frequencies in terms of detunings will prove particularly convenient, as it removes the need to consider a specific resonant frequency, and we will find that the most interesting behaviour occurs when $|y|$ is of order unity.

Equations (\ref{eqn:op_point_equation}) and (\ref{eqn:u0}) must be solved self-consistently to find the values of $v_0$ and $U_0$ for given applied power, which we will call the operating point of the device.
Multiple solutions may exist, in which case the realised operating point will depend on how the pump was brought to its final state, e.g. whether it was simply switched-on, or retuned from a higher or lower frequency.
The behaviour of nonlinear resonators driven by high-level monochromatic tones has been considered in detail previously\,\cite{thomas2020nonlinear}.

It is useful to define some new variables relating to the steady state behaviour that will simplify the subsequent analysis.
Let
\begin{equation}\label{eqn:def_p}
	p = 1 + 2 i \{ y_p - g(v_0) \},
\end{equation}
\begin{equation}\label{eqn:def_a_star}
	a_* = \frac{2 Q_r^2 |a_0|^2}{\pi \fres Q_c U_*}
	= \frac{2 Q_r^2 P_0 |\mathbf{k}^T \cdot \mathbf{a}_0|^2}
		{\pi \fres Q_c U_* |\mathbf{k}|^2 |\mathbf{a}_0|^2}
\end{equation}
and $\qr = Q_r / Q_c$.
The new variable $p$ is a dimensionless parameter that will prove more convenient for characterising the operating point of the device than $U_0$ or $v_0$ directly.
The second variable $a_*$ is a normalised form of the readout power relative to some chosen scale energy $U_*$.
In the rightmost part of (\ref{eqn:def_a_star}), we have re-expressed $|a_0|^2$ in terms of the total applied pump power $P_0 = |\mathbf{a}_0|^2$, with the factor of $|\mathbf{k}^T \cdot \mathbf{a}_0|^2 / (|\mathbf{k}|^2 |\mathbf{a}_0|^2)$ representing the fraction of this power contained in the active degree of freedom of the input signal.
The third variable is simply $Q_r$ normalised to $Q_c$.
However, since $Q_r^{-1} = Q_c^{-1} + Q_i^{-1}$ we have the convenient result $0 \leq \qr \leq 1$ in general and $Q_i \lessgtr Q_c$ for $\qr \lessgtr 0.5$.
Using these definitions in (\ref{eqn:u0}) and (\ref{eqn:gamma_p}), we find
\begin{equation}\label{eqn:gamma_p_norm}
	\Gamma_p = 1 - \frac{2 \qr}{p}
\end{equation}
and
\begin{equation}\label{eqn:U0_norm}
	\frac{U_0}{U_*} = \frac{a_*}{|p|^2},
\end{equation}
for later use.

The scattering matrix $\mathsf{S}_p$ can be measured experimentally with a vector network analyser (VNA) that has controllable readout power.
The only condition is that the frequency/power must be swept slowly enough to allow the internal state $v$ to relax to its steady-state value at every point recorded.
It is then possible to obtain $\Gamma_p$ from these measurements using (\ref{eqn:sp}); note that a single element of $\mathsf{S}_p$ is sufficient for this purpose, provided it is affected by $\Gamma_p$.
This in turn allows $p$ to be recovered, via (\ref{eqn:gamma_p_norm}).

\subsection{Small signal analysis}\label{sec:small_signal_analysis}

Assume that a signal, $\mathbf{a}_s (t)$, is applied in addition to the pump:
\begin{equation}
	\mathbf{a}(t) = \mathbf{a}_0 e^{2 \pi i \fpmp t} + \mathbf{a}_s (t).
\end{equation}
The active input component then becomes
\begin{equation}
	a(t) = a_0 e^{2 \pi i \fpmp t} + a_s (t),
\end{equation}
for $a_s(t) = \xvec^\dagger \cdot \mathbf{a}_s(t)$.
We will assume the signal is `small' in the sense $|a_s(t)| \ll |a_0|$ for all $t$.
Under this condition, we would expect all other variables to only differ slightly from their pump-only values.
Approximate
\begin{equation}\label{eqn:perturbed_v}
	v(t) = v_0 + v_s (t),
\end{equation}
\begin{equation}\label{eqn:perturbed_u}
	u(t) = u_0 e^{2 \pi i \fpmp t} + u_s (t)
\end{equation}
and
\begin{equation}\label{eqn:perturbed_b}
	b(t) = b_0 e^{2 \pi i \fpmp t} + b_s (t)
\end{equation}
where $v_0$, $u_0$ and $b_0$ were defined in the previous section, and subscript $s$ denotes the perturbation caused by the signal.
We are now able find a new set of dynamical equations describing the relationships between the perturbations.

Consider the behaviour of the state variable $v$ characterising the internal process driving the nonlinearity.
$h(v, U)$ in (\ref{eqn:v_state_equation}) can be expanded in a Taylor series about $(v_0, U_0)$ for (\ref{eqn:perturbed_v}) and (\ref{eqn:perturbed_u}) as follows:
\begin{widetext}
\begin{equation}\label{eqn:h_expansion}
\begin{aligned}
	h(v, U)
	&= h(v_0, U_0) + \left( \frac{\partial h}{\partial v} \right)_\op v_s (t)
	+ \left( \frac{\partial h}{\partial U} \right)_\op
		\left( \frac{\partial U}{\partial u} \right)_\op
		u_s (t)
	+ \left( \frac{\partial h}{\partial U} \right)_\op
		\left( \frac{\partial U}{\partial u^*} \right)_\op
		u^*_s (t) + \dots \\
	&= h(v_0, U_0) + \left( \frac{\partial h}{\partial v} \right)_\op v_s (t)
	+  2 \left( \frac{\partial h}{\partial U} \right)_\op
		\Re[ u_0^* e^{-2 \pi i \fpmp t} u_s(t) ] + \dots
\end{aligned}
\end{equation}
\end{widetext}
where the subscript $\op$ indicates that the derivatives are evaluated at the operating point: $v = v_0$ and $U = U_0$.
Substituting (\ref{eqn:perturbed_v}) and (\ref{eqn:h_expansion}) into (\ref{eqn:v_state_equation}), and noting that the contributions from the pump must cancel,
\begin{equation}\label{eqn:dvs_dt_1}
	\frac{dv_s}{dt} = \left( \frac{\partial h}{\partial v} \right)_\op v_s (t)
	+  2 \left( \frac{\partial h}{\partial U} \right)_\op
		\Re[ u_0^* e^{-2 \pi i \fpmp t} u_s(t) ],
\end{equation}
where we have retained only the terms to first-order in $u_s (t)$ and $v_s(t)$, and assumed that the higher order terms are small enough to ignore.
In this form, the derivative of $h$ with respect to $v$ defines an effective relaxation time $\tau_v$ for $v_s$ according to
\begin{equation}\label{eqn:def_tv}
	\frac{1}{\tau_v} = -\left( \frac{\partial h}{\partial v} \right)_\op.
\end{equation}
Similarly, the signal component of the mode amplitude, $u_s(t)$, acts as a forcing term for $v_s (t)$, with the derivative of $h$ with respect to $U$ determining the scaling factor.
Rather than using the derivative directly, it will be more convenient to work with the dimensionless scaling factor
\begin{equation}\label{eqn:def_alpha}
	\alpha = \frac{\tau_v U_0}{v_0} \left( \frac{\partial h}{\partial U} \right)_\op.
\end{equation}
Using (\ref{eqn:def_tv}) and (\ref{eqn:def_alpha}), we can write (\ref{eqn:dvs_dt_1}) in a more physically meaningful way:
\begin{equation}\label{eqn:dvs_dt_2}
	\frac{dv_s}{dt} = -\frac{1}{\tau_v} v_s (t)
	+  \frac{2 \alpha v_0}{\tau_v U_0}
		\Re[ u_0^* e^{-2 \pi i \fpmp t} u_s(t) ].
\end{equation}

An identical process can be carried out for the mode amplitude of the resonator.
To do so, expand $g(v)$ as a Taylor series around $v_0$ for (\ref{eqn:perturbed_v}), which gives
\begin{equation}\label{eqn:g_expansion}
	g(v) = g(v_0)
	+ \frac{\beta}{v_0} v_s(t)
	+ \frac{1}{2} \left( \frac{d^2 g}{d v^2} \right)_\op v_s (t)^2
	+ \dots
\end{equation}
where a second dimensionless scaling factor $\beta$ has been defined:
\begin{equation}\label{eqn:def_beta}
	\beta = v_0 \left( \frac{d g}{d v} \right)_\op.
\end{equation}

Substituting (\ref{eqn:perturbed_u}) and (\ref{eqn:g_expansion}) into (\ref{eqn:scalar_state_equation}), cancelling the pump terms, and again only keeping terms that are first order in $u_s$ and $v_s$, gives
\begin{equation}\label{eqn:dus_dt}
\begin{aligned}
	\frac{du_s}{dt} &=
	2 \pi i \fres \biggl\{ 1 + \frac{i}{2 Q_r} + \frac{g(v_0)}{Q_r} \biggr\} u_s(t) + \\
	& \frac{2 \pi i \beta \fres u_0}{v_0 Q_r} e^{2 \pi i \fpmp t} v_s(t)
	+ i \kappa a_s(t).
\end{aligned}
\end{equation}
In the case of a purely linear device, the principle of superposition holds, and (\ref{eqn:dus_dt}) would be the same as for the pump.
In (\ref{eqn:dus_dt}) we see two key differences as a result of the nonlinearity.
First, the term $g(v_0)/Q_r$ means the resonator behaves with resonant frequency and Q-factor determined by the pump, rather than the zero-power values.
Second, there is an additional source term, proportional to $e^{2 \pi i \fpmp t} v_s(t)$.
Although we could carry both state parameters,  $u_s$ and $v_s$, forward in the analysis, it is simpler to use (\ref{eqn:dvs_dt_2}) to eliminate $v_s$ in (\ref{eqn:dus_dt}), and solve purely for the mode amplitude, since only the latter directly affects the output.

Under the assumption that the nonlinearity state parameter, $v_s$, is zero at some time sufficiently far in the past, (\ref{eqn:dvs_dt_2}) has the solution
\begin{equation}\label{eqn:v1_solution}
	v_s (t) = \frac{2 \alpha v_0}{\tau_v U_0} \int_{-\infty}^t
	\Re \left[ u_0^* e^{-2 \pi i \fpmp t'} u_s (t') \right]
	e^{-(t - t') / \tau_v}
	\, dt'.
\end{equation}
We see that $v_s(t)$ has the form of a mixed signal: $u_s(t)$ multiplied by $e^{-2 \pi i \fpmp t}$.
In the limit where the effective response time $\tau_v$ is zero,  $v_s(t)$ traces the mixed signal exactly, but for finite $\tau_v$ the response is low-pass filtered down to a single-sided baseband of $\approx 2 \pi / \tau$.
Thus the nonlinear state parameter has frequency components at the difference between the pump and signal, which are generally slow.
The low-pass filtered spectral content of $v_s$ is then mirrored in the resonant frequency and Q-factor.

Substituting (\ref{eqn:v1_solution}) into (\ref{eqn:dus_dt}) gives
\begin{equation}\label{eqn:dus_dt_final}
\begin{aligned}
	&\frac{du_s}{dt} =
	2 \pi i \fres \biggl\{ 1 + \frac{i}{2 Q_r} + \frac{g(v_0)}{Q_r} \biggr\} u_s(t) + i \kappa a_s(t) + \\
	&\frac{4 \pi i \alpha \beta \fres u_0}{\tau_v U_0 Q_r}  e^{2 \pi i \fpmp t}
	\! \! \int_{-\infty}^t \! \! \!
	\Re \left[ u_0^* e^{-2 \pi i \fpmp t'} u_s (t') \right]
	e^{-(t - t') / \tau_v}
	\, dt',
\end{aligned}
\end{equation}
which is the governing equation for the signal part of the mode amplitude.
The last term on the right of (\ref{eqn:dus_dt_final}), which characterises the effect of the nonlinear modulation on the resonator's parameters, will prove to be the key to providing amplification.
Although explicit reference to the nonlinear state parameter has been removed, its role as an intermediate step is still apparent through the low-pass filtering.

(\ref{eqn:dus_dt_final}) allows the small-signal behaviour of the mode-amplitude to be calculated, but to complete the analysis we need a set of equations to relate it to the output signal.
These follow by substituting (\ref{eqn:perturbed_b}) into (\ref{eqn:scalar_output_equation}) and (\ref{eqn:b_vec_from_b}) and cancelling the pump terms, which gives
\begin{equation}\label{eqn:signal_scalar_output_equation}
	b_s(t) = a_s(t) + i \kappa u_s (t)
\end{equation}
and
\begin{equation}\label{eqn:signal_b_vec_from_b}
	\mathbf{b}_s(t) = \mathsf{S}_0 \cdot \mathbf{a}'_s (t) + b_s (t) \xvec^*.
\end{equation}

\subsection{Response to a sinusoidal signal}\label{sinusoidal_signal}

Consider the response when the applied signal is a single tone:
\begin{equation}
	\mathbf{a}_s (t) = \mathbf{a}_s e^{2 \pi i \fsgnl t},
\end{equation}
where the signal frequency is different to the pump frequency, $\fsgnl\neq f_{p}$.
To make progress, it is convenient to consider an intermediate function $\psi(t)$ defined by
\begin{equation}\label{eqn:def_psi}
	u_s (t) = \frac{\kappa Q_r}{\pi \fres} \psi (t) e^{2 \pi i \fpmp t}.
\end{equation}
Noting that
\begin{equation}
	e^{2 \pi i \fpmp t} \frac{d}{dt} \left\{ e^{-2 \pi i \fpmp t} u_s(t) \right\}
	= -2 \pi i  \fpmp u_s(t) + \frac{du_s}{dt}
\end{equation}
and expanding out the real part $\Re$, we can rewrite (\ref{eqn:dus_dt_final}) in terms of $\psi(t)$ as
\begin{equation}\label{eqn:dpsi_equation}
\begin{aligned}
	\frac{Q_r}{\pi \fres} \frac{d\psi}{dt} &=
	- p \psi(t) - \frac{q}{\tau_v} \int_{-\infty}^t \psi (t') e^{-(t - t') / \tau_v} \, dt' - \\
	&\frac{q}{\tau_v} e^{2 i \phi_p} \int_{-\infty}^t \psi^*(t') e^{-(t - t') / \tau_v} \, dt'
	+ i a_s e^{2 \pi i (\fsgnl - \fpmp) t}
\end{aligned}
\end{equation}
where $p$ is as defined in (\ref{eqn:def_p}), $e^{2 i \phi_p} = u_0^2 / U_0$ is a pure phase-factor and $q$ is a new dimensionless factor defined by
\begin{equation}\label{eqn:def_q}
	q = -2 i \alpha \beta.
\end{equation}
We will refer to $q$ as the \emph{modulation factor}; its physical significance will be discussed in more detail in Section \ref{sec:modulation_factor}.

The drive term and presence of $\psi^*$ in (\ref{eqn:dpsi_equation}) together imply that $\psi(t)$ must have a components at frequencies $\fsgnl - \fpmp$ and $-(\fsgnl - \fpmp)$, with the latter corresponding to the idler frequency in the final signal.
As such, we consider a solution of the form
\begin{equation}\label{eqn:psi_trial}
	\psi (t) = c_s e^{2 \pi i(\fsgnl - \fpmp) t}
		+ e^{2 i \phi_p} c^*_i  e^{-2 \pi i (\fsgnl - \fpmp) t}.
\end{equation}
After substituting (\ref{eqn:psi_trial}) into (\ref{eqn:dpsi_equation}) and evaluating the integrals, decomposing the result into components at the two frequencies yields a pair of coupled linear equations for $c_i$ and $c_s$.
The latter can be expressed in matrix form as
\begin{equation}\label{eqn:cs_and_ci_equation}
	\Biggl\{ \left( \begin{matrix}
		p + x  & 0 \\
		0 & p^* + x
	\end{matrix} \right)
	+ \frac{1}{1 + r x} \left( \begin{matrix}
		q  & q \\
		q^* & q^*
	\end{matrix} \right)
	\Biggr\} \cdot
	\left( \begin{matrix}
		c_s \\
		c_i
	\end{matrix} \right) \\
	=
	\left( \begin{matrix}
		i a_s \\
		0
	\end{matrix} \right),
\end{equation}
where
\begin{equation}\label{eqn:def_x}
	x = 2 i (y_s - y_p),
\end{equation}
$y_s = Q_r (\fsgnl - \fres) / \fres$ is the detuning of the signal frequencies in linewidths and
\begin{equation}\label{eqn:def_r}
	r = \frac{\pi \fres \tau_v}{Q_r}
\end{equation}
is the ratio of time of response time of nonlinearity state parameter $v$, at the operating point, to the ring-down time of the resonator.
The parameter $r$ provides a dimensionless measure of the significance of the relaxation time of $v$ to the amplifier's performance: the larger $r$, the greater the effect.
(\ref{eqn:cs_and_ci_equation}) can be solved by matrix inversion, yielding
\begin{equation}\label{eqn:cs_and_ci}
	\left( \begin{matrix}
		c_s \\ c_i
	\end{matrix} \right)
	= \frac{i a_s}{K(x)}
	\left( \begin{matrix}
		(p^* + x) (1 + rx) + q^* \\
		-q^*
	\end{matrix} \right)
\end{equation}
for
\begin{equation}\label{eqn:def_big_K}
	K(x) = (p + x) (p^* + x) (1 + r x) + (p + x) q^* + (p^* + x) q.
\end{equation}

The active component of the output signal must also contain the two frequencies.
Let
\begin{equation}
	b_s (t) = b_s e^{2 \pi i \fsgnl t} + b_i e^{2 \pi i (2 \fpmp - \fsgnl)t}
\end{equation}
and define complex reflection coefficients $\Gamma_s = b_s / a_s$ and $\Gamma_i = b_i / a_s^* $ as measured for the signal and idler frequencies respectively.
Some care is needed in interpreting $\Gamma_i$, as the reflection is between different frequency components.
Note also the conjugation of $a_s$ in the definition of $\Gamma_i$.
Using (\ref{eqn:cs_and_ci}), (\ref{eqn:psi_trial}) and (\ref{eqn:def_psi}) together to substitute for $u_s(t)$ in (\ref{eqn:signal_scalar_output_equation}), after identifying the relevant frequency components we find
\begin{equation}\label{eqn:signal_reflection_coefficient}
	\Gamma_s = 1 - \frac{2 \qr \{ (p^* + x) (1 + r x) + q^* \}}{K(x)}
\end{equation}
and
\begin{equation}\label{eqn:idler_reflection_coefficient}
	\Gamma_i = -\frac{2 \qr q}{K^*(x)} e^{2 i \phi _p}.
\end{equation}
These results fully characterise the scattering behaviour for the active field components at the signal and idler frequencies.

It then follows from (\ref{eqn:signal_b_vec_from_b}) that the full scattering matrix $\mathsf{S}_s$ of the circuit at the signal frequencies, as defined by $\mathbf{b}_s = \mathsf{S}_s \cdot \mathbf{a}_s$, is given by
\begin{equation}\label{eqn:signal_scattering_matrix}
	\mathsf{S}_s = \mathsf{S}_0 - (1 - \Gamma_s) \xvec^* \xvec^\dagger.
\end{equation}
In the case of the idler frequency, we must define the scattering matrix by $\mathbf{b}_i = \mathsf{S}_i \cdot \mathbf{a}_s^*$ in analogy with $\Gamma_i$.
There is no contribution from $\mathbf{a}'_s$ in (\ref{eqn:signal_b_vec_from_b}) at idler frequencies, so we have
\begin{equation}\label{eqn:idler_scattering_matrix}
	\mathsf{S}_i = \Gamma_i \, \xvec^* \xvec^\dagger.
\end{equation}

(\ref{eqn:signal_reflection_coefficient})--(\ref{eqn:idler_scattering_matrix}) allow for power gain between the input and output signals under certain conditions, allowing the resonator to be used as an amplifier.
The exact conditions and resulting behaviour are discussed in detail in Section \ref{sec:amplification_regimes}.

\subsection{The modulation factor}\label{sec:modulation_factor}

Previously, we introduced the dimensionless modulation factor $q$, defined by (\ref{eqn:def_q}), and now we consider its physical interpretation.
Using (\ref{eqn:def_alpha}), (\ref{eqn:def_beta}) and (\ref{eqn:def_tv}), it follows from (\ref{eqn:def_q}) that
\begin{equation}\label{eqn:alt_def_q}
	q = 2 i U_0  \frac{
		\bigl( \tfrac{\partial g }{\partial v} \bigr)_\op
		\bigl( \tfrac{\partial h }{\partial U} \bigr)_\op
	}{
		\bigl( \tfrac{\partial h }{\partial v} \bigr)_\op
	},
\end{equation}
where again we use the subscript $\op$ to indicate evaluation of the derivative at the operating point.
To appreciate the significance of this combination, consider what happens when the pump is changed in some differential way; either in frequency or amplitude.
The steady-state operating point moves from $(U_0, v_0)$ to $(U_0 + \delta U, v_0 + \delta v)$.
(\ref{eqn:op_point_equation}) must be satisfied at the new operating point, and so
\begin{equation}
	0 = h (U_0 + \delta U, v_0 + \delta v)
	\approx \left( \frac{\partial h }{\partial U} \right)_\op \delta U
	+ \left( \frac{\partial h }{\partial v} \right)_\op \delta v
\end{equation}
to first order, since $h(U_0, v_0)=0$ by definition.
Hence
\begin{equation}
	\delta v = -\frac{
		\bigl( \tfrac{\partial h }{\partial U} \bigr)_\op
	}{
		\bigl( \tfrac{\partial h }{\partial v} \bigr)_\op
	} \delta U,
\end{equation}
and the change in the resonant frequency $\delta' \! \fres$, and internal dissipation factor $\delta' Q_i^{-1}$, with respect to the original operating point, are found to be
\begin{equation}\label{eqn:pole_shift}
\begin{aligned}
	&\frac{\delta' \! \fres}{\fres} + \frac{i \delta' Q_i^{-1}}{2}
	\approx \frac{1}{Q_r} \left( \frac{\partial g}{\partial v} \right)_\op
	\delta \nu \\
	&\approx
	- \frac{U_0}{Q_r} \frac{
		\bigl( \tfrac{\partial g }{\partial v} \bigr)_\op
		\bigl( \tfrac{\partial h }{\partial U} \bigr)_\op
	}{
		 \bigl( \tfrac{\partial h }{\partial v} \bigr)_\op
	} \frac{\delta U}{U_0} \approx \frac{i q}{2 Q_r} \frac{\delta U}{U_0}
\end{aligned}
\end{equation}
using (\ref{eqn:alt_def_q}).
We have used $\delta'$ here to distinguish these changes in $f_0$ and $Q_i$, measured relative to the operating point values, from those in (\ref{eqn:model_non_linearity}), which are measured relative to the zero readout power values.
(\ref{eqn:pole_shift}) shows we may identify $q$ as the rate of change of the complex pole of the resonator with respect to fractional changes in the stored energy, as measured at the operating point, while maintaining the steady state condition (\ref{eqn:op_point_equation}).
This definition will prove useful for the experimental measurement of $q$.

\subsection{Obtaining operating point parameters from measurements}

(\ref{eqn:signal_reflection_coefficient})--(\ref{eqn:idler_scattering_matrix}) are parametrised in terms of the operating point parameters $p$, $q$ and $r$.
An important question, therefore, is how can these quantities be found experimentally?

We have already discussed the measurement of $p$ at the end of Section \ref{sec:response_to_pump} and similar steady state measurements can also be used to recover the modulation factor $q$.
The key is to measure the change $\delta \Gamma_p$ in $\Gamma_p$ after an infinitesimal change $\delta P_0$ has been made in the applied power and then the device allowed to settle to its new operating point.
Under these conditions we would expect the resonant frequency, internal dissipation factor, total energy stored and state parameter $p$ to also change by infinitesimal amounts $\delta' \! \fres$, $\delta' Q_i^{-1}$, $\delta U$ and $\delta p$, respectively.
The following set of relationships between these infinitesimal quantities can be found by forming the differentials of $p$, $U$ and $\Gamma_p$ from (\ref{eqn:def_p}), (\ref{eqn:gamma_p_norm}) and (\ref{eqn:U0_norm}):
\begin{equation}\label{eqn:differential_p}
	\delta p = -2 i \delta g
	= - \frac{2 i}{Q_r} \left\{
		\frac{\delta' \! \fres}{\fres} + \frac{i \delta' Q_i^{-1}}{2}
	\right\}
\end{equation}
\begin{equation}\label{eqn:differential_U}
	\frac{\delta U}{U_0} = \frac{\delta a_*}{a_*} - 2 \Re \left[ \frac{\delta p}{p} \right]
\end{equation}
\begin{equation}\label{eqn:differential_gamma_p}
	\delta \Gamma_p
	= (1 - \Gamma_p) \frac{\delta p}{p}
	= \frac{1}{2 \qr} (1 - \Gamma_p)^2 \delta p.
\end{equation}
Here $\delta a_*$ is the infinitesimal change in the normalised readout power, which is related to the change in readout power by $\delta a_* / a_* = \delta P_0 / P_0$.
(\ref{eqn:differential_gamma_p}) can be used to substitute for $\delta p$ in the other two expressions to give
\begin{equation}\label{eqn:differential_f_and_q_shift}
	\frac{\delta' \! \fres}{\fres} + \frac{i \delta' Q_i^{-1}}{2}
	= \frac{i \delta \Gamma_p}{(1 - \Gamma_p)^2 Q_c}
\end{equation}
and
\begin{equation}\label{eqn:fractional_U}
	\frac{\delta U}{U_0}
	= \frac{\delta P}{P_0} - 2 \Re \left[ \frac{\delta \Gamma_p}{1 - \Gamma_p} \right]
\end{equation}
Hence it follows from (\ref{eqn:pole_shift}) that $q$ can be calculated from $\Gamma_p$ and $\delta P_0$ using
\begin{equation}\label{eqn:q_from_measurements}
	q = \frac{2 \qr \delta \Gamma_p}{(1 - \Gamma_p)^2}
	\left\{ \frac{\delta P_0}{P_0} - 2 \Re \left[ \frac{\delta \Gamma_p}{1 - \Gamma_p} \right]
	\right\}^{-1}.
\end{equation}

The remaining parameter $r$ characterises the relaxation time of the internal variable that drives the parametric changes, and so we would not expect it to be recoverable from steady-state measurements.
However, we will see in the next section that steady state measurements of $p$ and $q$ allow the bandwidth of the amplifier in the limit $r \rightarrow \infty$ to be determined.
The ratio of this value and the measured bandwidth can then be used to determine $r$.

\section{High-gain behaviour}\label{sec:amplification_regimes}

Consider the conditions on $\fsgnl$, $p$ and $q$ under which the pumped resonator can provide appreciable gain for a single tone.
It will be sufficient to consider the behaviour of the active signal components, as characterised by the signal and idler reflection coefficients $\Gamma_s$ and $\Gamma_p$, as only these components can be amplified.

One approach would be to carry out a full pole-zero analysis on (\ref{eqn:signal_reflection_coefficient}) and (\ref{eqn:idler_reflection_coefficient}).
Since the highest order polynomial appearing in both expressions is cubic, more specifically $K(x)$, it is possible to write down general analytic expressions for the poles and zeros.
However, the algebraic complexity of these expressions obscures the underlying physics.
Instead we focus on the behaviour in the high-gain regions, where the expressions for the poles and zeros have simple forms.

A particular strength of our formalism is that it takes account of the possibly finite response time of the nonlinear mechanism.
A convenient measure of the operating regime of the device with respect to the speed of the nonlinearity is the dimensionless parameter $r$, as defined in (\ref{eqn:def_r}) as the ratio of $\tau_v$, the linearised response time of the nonlinear mechanism at the operating point, to the ringdown time $\tau_r = Q_r / \pi \fres$ of the resonator.
The limit $r=0$ corresponds to the resonator's characteristics changing instantaneously with stored energy.
Here, or more generally when $r \ll 1$, the overall response time is dominated by resonator's electrical behaviour.
This regime is assumed, often implicitly, at the outset of most published analyses.
Here we can also consider the case $r > 1$, where the behaviour of the amplifier is limited by the rate at which the resonator's parameters can change.
Below, we start by analysing the simpler case $r=0$ to determine the conditions for high gain, and then extend the analysis to determine how the behaviour is affected by finite $r$.

Before preceding it is useful to highlight a number of constraints on $x$, $p$ and $q$.
First, it follows from (\ref{eqn:def_x}) that $x = 2 i Q_r (\fsgnl - \fpmp) / \fres$, and is therefore purely imaginary.
Second, under the assumption that the parametric response of the resonator can only ever be passive, diminishes with time when the excitation is removed, then the power dissipated can only ever be reduced to zero at best.
This imposes the restriction $\delta Q_i^{-1} \geq - Q_i^{-1}$, which through
(\ref{eqn:model_non_linearity}) means that we must have $\Im [g(v)] \geq - Q_r / 2Q_i \geq -1/2$ for all $v$ as $Q_r \leq Q_i$ by definition.
Applying this inequality to (\ref{eqn:def_p}) and noting that $q$ is the first-order change in $p$ for the change $\delta U$ induced in $U_0$, we find
\begin{equation}\label{eqn:re_p_plus_q}
	\Re[p], \Re[p + q] \geq 0
\end{equation}
at all operating points.

\subsection{Instantaneous nonlinearity}\label{sec:amplification_zero_r}

\begin{figure}
\centering
\includegraphics[width=8cm]{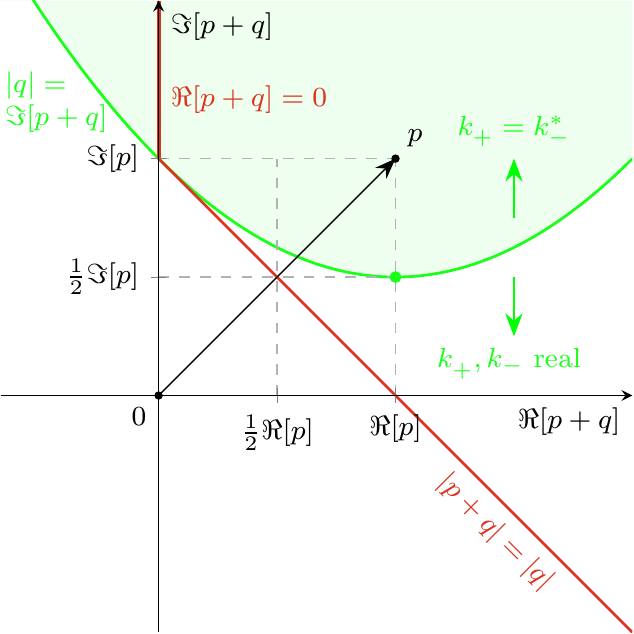}
\caption{\label{fig:p_plus_q}
Loci of $p+q$ in the Argand plane in different operating regimes for given $p$.
(\ref{eqn:re_p_plus_q}) means $p$ and $p + q$ are constrained to lie to the right of the origin.
The green line shows the values of $p + q$ for which $|q| = |\Im[p+q]|$; for values of $p + q$ lying above this line it the case that $k_+ = k^*_-$, whereas for values lying below it $k_+$ and $k_-$ are purely real.
The red line shows the values of $p + q$ for which the signal gain is infinite.
The section lying along the imaginary axis is where $\Re[p + q] = 0$, whereas the sloping section is where $|p + q| = |q|$.
}
\end{figure}

The expressions for $\Gamma_s$ and $\Gamma_i$ simplify significantly when $r=0$.
$K(x)$, defined in (\ref{eqn:def_big_K}), reduces to the quadratic
\begin{equation}\label{eqn:kappa_r_zero}
	K (x) = x^2 + 2 \Re[p + q] x + | p + q|^2 - |q|^2,
\end{equation}
which can be factored as
\begin{equation}\label{eqn:kappa_r_zero_factored}
	K (x) = (k_+ + x) (k_- + x),
\end{equation}
where
\begin{equation}\label{eqn:def_kpm}
\begin{aligned}
	k_\pm &=  \Re[p + q] \pm \sqrt{\Re[p + q]^2 + |q|^2 - |p + q|^2} \\
	&=  \Re[p + q] \pm \sqrt{|q|^2 - \Im[p + q]^2}.
\end{aligned}
\end{equation}
After substituting (\ref{eqn:kappa_r_zero_factored}) into (\ref{eqn:signal_reflection_coefficient}) and (\ref{eqn:idler_reflection_coefficient}) with $r$ set to zero, we can now carry out a partial fraction decomposition in both expressions, yielding
\begin{equation}\label{eqn:full_gs_r_zero}
	\Gamma_s (x) = 1 - \frac{2 \qr}{(k_+ - k_-)} \left\{
		\frac{k_+ - (p + q)^*}{k_+ + x} -
		\frac{k_- - (p + q)^*}{k_- + x}
	\right\}
\end{equation}
and
\begin{equation}\label{eqn:full_gi_r_zero}
	\Gamma_i (x) = -\frac{2 \qr}{(k_+ - k_-)^*} \left\{
		\frac{q}{(k_+ + x)^*} -
		\frac{q}{(k_- + x)^*}
	\right\} e^{2 i \phi_p}.
\end{equation}

Useful amplification can be achieved when either the signal or idler gains are large: $|\Gamma_s (x)|^2 \gg 1$ or $|\Gamma_i (x)|^2 \gg 1$ for some value of $x$, where we remind the reader the latter is the scaled detuning of the signal from the pump frequency.
From (\ref{eqn:full_gs_r_zero}) and (\ref{eqn:full_gi_r_zero}), we see this requires the existence of ranges of $x$ for which $|k_+ + x|$ or $|k_- + x|$ is sufficiently small, and in what follows we will consider the conditions on $p$ and $q$ for such ranges to exist.
In reality, we cannot control $p$ and $q$ independently, or indeed set them to specific values.
However later, when considering a specific nonlinearity, we will be able to translate these conditions onto constraints on the parameters we can control, such as pump-frequency and readout power.

The sign of the radicand $|q|^2 - \Im[p+q]^2$ in (\ref{eqn:def_kpm}) distinguishes two different regimes for the values of $k_+$ and $k_-$ depending on the values of $p$ and $q$.
The threshold between them is indicated by
\begin{equation}\label{eqn:k_case_threshold}
	|q|^2 - \Im[ p + q]^2 = 0
\end{equation}
which can be rearranged to
\begin{equation}\label{eqn:k_case_threshold_parabola}
	\Im[p + q] = \frac{(\Re[p + q] - \Re[p])^2}{2 \Im[p]}
	+ \frac{\Im[p]}{2}.
\end{equation}
This defines a parabolic locus in the Argand plane of $p + q$ for known $p$, as shown by the green line in Figure \ref{fig:p_plus_q}.
For $p + q$ in the region below this line, the radicand is positive, and so $k_+$ and $k_-$ are both real.
For $p + q$ in the region above the line, the radicand is negative, and so $k_+$ and $k_-$ are complex, and $k_+ = k_-^*$.

We will now identify four possible cases, I--IV, where the signal gain can in principle become large.
Cases I and II occur for the regime where $k_+$ and $k_-$ are both real.
In this regime, $|k_+ + x|$ and $|k_+ + x|$ are both minimized at $x = 0$, i.e. when the signal frequency is the same as the pump frequency.
The gain near this point will then be large if either $|k_+|$ or $|k_-|$ is sufficiently small, which we identify as Cases I and II respectively.
Cases III and IV occur in the regime in which $k_+$ and $k_-$ are both complex.
In this regime, $|k_+ + x|$ achieves its minimum value of $|\Re[k_+]|$ at $x = - i \Im[k_+]$.
The gain near this point will therefore be large if $|\Re[k_+]|$ is sufficiently small and we call this Case III.
Similarly, the same analysis for $|k_- + x|$ indicates the gain will be large near $x=- i \Im[k_-]$ when $|\Re[k_-]|$ is sufficiently small and we call this Case IV.

Each of the cases introduced comprises a requirement on $x$ and a condition that a target variable is sufficiently small.
To identify values for $p$ and $q$ for which each case is possible, we will consider the limit case where target variable is zero.
Strictly the gain is infinite in this limit and the values we find will turn out to be those at the switching points of the device between possible operating points.
However, by operating with $p$ and $q$ near these values large, but finite, gain can be achieved.

In Case I, we therefore look for values of $p$ and $q$ for which $|k_+| = 0$ for $|q|^2 - \Im[ p + q]^2 > 0$.
Using (\ref{eqn:def_kpm}) and (\ref{eqn:re_p_plus_q}), we see this is only the case if both $\Re[p + q] = 0$ and $|q| = \Im[p + q]$ together.
We will shortly describe why these conditions are difficult to achieve in practice.
In Cases III and IV, we look for the values of $p$ and $q$ for which $|\Re[k_+]| = 0$ and $|\Re[k_-]| =0$ respectively, with $|q|^2 - \Im[ p + q]^2 < 0$ in both cases.
It follows straightforwardly from (\ref{eqn:def_kpm}) that we require $\Re[p + q] =0$ in both cases.
The locus of values of $p+q$ satisfying this condition are shown the vertical section of the red line in Figure \ref{fig:p_plus_q}.
Finally, in Case II we need $|k_-| = 0$ for $|q|^2 - \Im[ p + q]^2 > 0$, which is achieved when $|p+q|^2 = |q|^2$.
By writing $q$ as $(p + q) - p$, the latter condition can be rearranged into
\begin{equation}\label{eqn:gain_cond}
	\Im[p + q] = - \frac{\Re[p]}{\Im[p]} \Re[p + q] + \frac{|p|^2}{2 \Im [p]}.
\end{equation}
which shows the appropriate values of $p + q$ for given $p$ lie along a straight line in the Argand plane.
Equation (\ref{eqn:gain_cond}) corresponds to the diagonal section of the red line in Figure \ref{fig:p_plus_q}.
It can be proven algebraically that this section of the red line is tangential to the green line at their point of intersection.

Based on this discussion, we expect large gain for given $p$ whenever $p + q$ falls sufficiently close to the red line in Figure \ref{fig:p_plus_q}.
In Cases I, III and IV, significant gain is achieved only when $\Re[p + q] \rightarrow 0$.
This requirement is difficult to realise in practice because the power dissipated in the resonator must decrease to zero as the stored energy is increased.
Of the common nonlinear processes associated with superconducting resonators, dielectric two-level system (TLS) loss is the only mechanism that behaves in this manner.
However, even in this case, it is unlikely that the total power dissipated can be reduced to near-zero, because in a real device other loss mechanisms, such as quasiparticle dissipation, act simultaneously.
Nevertheless, a TLS-based parametric amplifier is an interesting possibility.
Despite this possibility, for the rest of the paper we will focus on operation of the device in Case II: $|p + q| \approx |q|$ and $|q| > | \Im [p + q]|$.
The latter is more readily achievable in a real device, e.g. with a purely reactive nonlinearity.

In Case II, the reflection coefficient takes a particularly simple form.
Assume the device is operated sufficiently close to the condition $|p + q| = |q|$ that $||q|^2 - |p+q|^2| \ll \Re[p + q]^2$.
It follows from (\ref{eqn:def_kpm}) that we may approximate
\begin{equation}\label{eqn:approx_km}
	k_- \approx \frac{|p + q|^2 - |q|^2}{2 \Re[p + q]}
\end{equation}
and
\begin{equation}\label{eqn:approx_kp}
	k_+ \approx 2 \Re[p + q].
\end{equation}
The approximation also implies $|k_-| \ll 1, k_+$.
Similarly,
\begin{equation}
\begin{gathered}
	k_+ - (p + q)^* \approx (p + q) \\
	k_- - (p + q)^* \approx -(p + q)^*.
\end{gathered}
\end{equation}
Hence the signal reflection coefficient, as given by (\ref{eqn:full_gs_r_zero}), is
\begin{equation}\label{eqn:approx_gs_r_zero}
\begin{aligned}
	\Gamma_s (x) &\approx 1 - \frac{2 \qr}{k_+ k_-} \left\{
		\frac{k_-}{k_+} \frac{p + q}{1 + x / k_+} +
		\frac{(p + q)^*}{1 + x /k_-}
	\right\} \\
	&\approx 1 - \frac{(p + q)^*}{\Re[p + q]}
	\frac{\qr}{k_- + x }.
\end{aligned}
\end{equation}
By definition, $|p + q| / \Re[p + q] > 1$ and the earlier condition $|k_-| \ll 1$ implies high gain at $x=0$.

It is useful at this point to reintroduce the physical signal frequencies using (\ref{eqn:def_x}).
We may then rewrite (\ref{eqn:approx_gs_r_zero}) as
\begin{equation}\label{eqn:approx_gs_r_zero_physical_var}
	\Gamma_s (\fsgnl) \approx 1 - \frac{\sqrt{G_s} e^{i \phi_s}}{1 + 2 i (\fsgnl - \fpmp) / \bwzero}
\end{equation}
where we may now identify
\begin{equation}\label{eqn:general_signal_power_gain}
	G_s = \frac{|p + q|^2 \qr^2}{\Re[p + q]^2 |k_-|^2}
	= \frac{4 |p + q|^2 Q_r^2}{(|p + q|^2 - |q|^2)^2 Q_c^2}
\end{equation}
as the maximum signal frequency power gain,
\begin{equation}\label{eqn:general_signal_phase_factor}
	e^{i \phi_s} = \frac{(p + q)^* |k_+||k_-|}{k_+ k_- |p + q| }
\end{equation}
as the phase shift when $\fsgnl = \fpmp$ and
\begin{equation}\label{eqn:general_signal_bandwidth_r_zero}
	\bwzero = \frac{k_- \fres}{Q_r}
	= \frac{\{|p + q|^2 - |q|^2\} \fres}{2 Q_r \Re[p + q]}
\end{equation}
as the 3\,dB bandwidth of the amplifier gain.
Strictly, the physical interpretations of these quantities apply only when $G_s$ is sufficiently large that the effect of the first term in (\ref{eqn:general_signal_power_gain}) can be ignored.
We see that the pumped resonator can function as a high-gain, phase-insensitive, reflection amplifier for a single tone, and that the corresponding gain has a single-pole response centred about the pump frequency.
In addition, the maximum power gain and bandwidth are interrelated via their mutual dependence on $k_-$; broadly speaking, the bandwidth is expected to be narrower the higher the gain.

\subsection{Slow nonlinearity}\label{sec:amplification_finite_r}

Now consider the effect of non-zero $r$, in which case the parametric process is slowed by the response time of the underlying physical mechanism driving the nonlinearity.
Notice that $r$ appears in the expressions for $\Gamma_s$ and $\Gamma_i$ scaled by $x$, i.e. as $rx$.
As a result, if we distinguish different regimes based on the value of $r x$ these will be achieved for different values of $x$ for given $r$.
In the regime where $r x \ll 1$, the expressions for signal and idler gain reduce to those of Section \ref{sec:amplification_zero_r}
This is a crucial observation, as it means for any $r$ there will always be some range of $x$ where the device behaves as for $r=0$ (even it is very small) and for which gain is possible if the conditions on $p$ and $q$ are met.

In the opposite limit, $rx \gg 1$, we expect that there will be sufficiently large $rx$ for which
\begin{equation}
	K (x) \approx (x + p) (x + p^*) rx
\end{equation}
and
\begin{equation}
	(x + p^*) (rx + 1) + q^* \approx (x + p^*) rx.
\end{equation}
Certainly, this requires $rx \gg 1$.
In this limit the signal gain becomes
\begin{equation}
	\Gamma_s \approx 1 - \frac{2 \qr}{p + x}
\end{equation}
Now, the limited response time nulls the dynamical nonlinear behaviour: $\Gamma_s$ traces out a resonance curve with the Q-factor and resonant frequency determined by the static power loading of the pump.
It is also possible to show that no idler signal will be seen ($\Gamma_i = 0$).
Essentially, the signal frequency is beyond the band edge imposed by the finite response time of the nonlinearity.
A key question is what happens between these two extremes?

To investigate such effects we consider a device operating in the high-gain region described in the previous section, where $|q| > |\Im[p + q]|$ and $|p + q | \approx |q|$ and we expect gain for small $x$ (Case II).
To do so it is convenient to write $K(x)$ in the form
\begin{equation}\label{eqn:kappa_kpm}
	K(x) = (x + k_+) (x + k_-) + (x + p) (x + p^*) r x
\end{equation}
from which we may derive the following first-order Taylor series approximation for $K(x)$ near $x=0$:
\begin{equation}\label{eqn:kappa_near_zero}
	K(x) \approx k_+ k_- + (k_+ + k_- + r |p|^2) x.
\end{equation}
Likewise, we may approximate
\begin{equation}\label{eqn:numerator_approx}
	(x + p^*)(rx + 1) + q^*
	\approx (p + q)^* + (r p^* + 1) x
\end{equation}
to first-order.
Where these approximations are valid, the signal gain simplifies to
\begin{equation}\label{eqn:approx_signal_gain}
	\Gamma_s (\fsgnl) \approx 1 - \frac{\sqrt{G_s} e^{i \phi_s}}
	{1 + 2 i (\fsgnl - \fpmp) / \bw},
\end{equation}
where $G_s$ and $\exp(i \phi_s)$ are as defined earlier and
\begin{equation}\label{eqn:general_signal_bandwidth}
	\bw = \frac{\bwzero}{1 + r|p|^2 / 2 \Re[ p + q]}
\end{equation}
is the modified bandwidth when $r$ is non-zero.
(\ref{eqn:general_signal_bandwidth}) reproduces (\ref{eqn:approx_gs_r_zero}) up to an $r$-dependent factor in the bandwidth.
Interestingly, in the limit of large $r$ the bandwidth does not limit to $1 / (2 \pi \tau_v)$, as might have been expected, but instead to
\begin{equation}
	\bw \approx \frac{2 \Re[p + q]}{|p|^2} \frac{\bwzero}{r},
\end{equation}
where $\bwzero$ bandwidth in the instantaneous limit, $r=0$.

The same analysis can also be repeated for the idler reflection coefficient as given by (\ref{eqn:idler_reflection_coefficient}), in which case it is found that
\begin{equation}\label{eqn:approx_idler_gain}
	\Gamma_i (\fsgnl) \approx
	-\frac{\sqrt{G_i} e^{i (\phi_i + 2 \phi_p)}}
	{1 - 2 i (\fsgnl - \fpmp) / \bw}.
\end{equation}
$\Delta f_{3\,\text{dB}}$ is as for the signal, (\ref{eqn:general_signal_bandwidth}), but
\begin{equation}\label{eqn:general_idler_power_gain}
	G_i = \frac{4 Q_r^2 |q|^2}{(|p + q|^2 - |q|^2)^2 Q_c^2}
\end{equation}
and
\begin{equation}
	e^{i \phi_i} = \frac{|k_+| |k_-| q}{k_+ k_- |q|}.
\end{equation}
Since the condition for high gain is $|p + q| \approx |q|$, the signal and idler gains are approximately equal.

To finish, consider the gain-bandwidth product.
Taking the product of (\ref{eqn:general_signal_power_gain}) and (\ref{eqn:general_signal_bandwidth}) gives
\begin{equation}\label{eqn:gain_band_prod}
	\sqrt{G_s} \bw \approx  \frac{2 f_0}{Q_c} \frac{|p+q|}{2 \Re[p + q] + r|p|^2},
\end{equation}
where the prefactor $2f_0 /Q_c$ is the gain bandwith product of the unpumped resonator.
Notice that the gain-bandwidth products falls when the parametric process slows the device.

(\ref{eqn:approx_signal_gain}) and (\ref{eqn:approx_idler_gain}) are key results of this paper.
They show that a pumped resonator that would be expected to produce gain when $r=0$ will still produce gain when $r$ is finite, i.e. when the parametric process is rate-limited.
The gain for signal and idler frequencies are approximately equal, and show a single-pole response response centred on the pump.
The achievable gain is not affected by $r$ to first-order.
Instead, it imposes a restriction on the bandwidth.

\subsection{Phase-sensitivity}\label{sec:phase_sensitivity}

To this point we have considered amplification of a single frequency tone.
In the case of a signal with more complicated spectral content, each spectral component will be amplified in the same way as a tone of the same frequency.
However, care must be taken if the spectral content spans the pump frequency.
In this case it is possible for the idler tones generated by frequency components on one side of the pump frequency to interfere with the amplification of those components on the other side.

Parametric amplification is often phase-sensitive, and in some way the behaviour just described is an example of this phenomena.
To see this, it is helpful to consider the amplification of a signal of the form
\begin{equation}\label{eqn:am_modulation}
	a_s (t) = a_m e^{i \phi_1} e^{2 \pi i \fsgnl t}
	+ a_m e^{-i \phi_2} e^{2 \pi i (2 \fpmp - \fsgnl) t}
\end{equation}
for real $a_m$, which comprises two tones of equal amplitude, but potentially different phase, evenly distributed around the pump frequency.
Assuming the high-gain regime of the preceding section and superposing the signals generated by the two spectral components individually, we expect
\begin{equation}
\begin{aligned}
	b_s(t) = &\left[ 1 - \frac{\sqrt{G_s} e^{i \phi_s}} {1 + 2 i (\fsgnl - \fpmp) / \bw} \right]
		a_m e^{i \phi_1} e^{2 \pi i \fsgnl t}  \\
	&-\frac{\sqrt{G_i} e^{i (\phi_i + 2 \phi_p)}}
	{1 - 2 i (\fsgnl - \fpmp) / \bw} a_m e^{-i \phi_1} e^{2 \pi i (2 \fpmp - \fsgnl) t} \\
	&+ \left[ 1 - \frac{\sqrt{G_s} e^{i \phi_s}} {1 - 2 i (\fsgnl - \fpmp) / \bw} \right]
		a_m e^{i \phi_2} e^{(2 \fpmp - \fsgnl) t} \\
	&- \frac{\sqrt{G_i} e^{i (\phi_i + 2 \phi_p)}}
	{1 + 2 i (\fsgnl - \fpmp) / \bw} a_m e^{-i \phi_2} e^{2 \pi i \fsgnl t}.
\end{aligned}
\end{equation}
Remembering that $G_i \approx G_s$ in the high gain regime and assuming $G_s \gg 1$, we may further approximate
\begin{widetext}
\begin{equation}
\begin{aligned}
	b_s(t)
	\approx &
	- \frac{2 \sqrt{G_s} e^{i (\phi_s + \phi_i + 2 \phi_p + \phi_1 - \phi_2) / 2} }
	{1 + 2 i (\fsgnl - \fpmp) / \bw}
	\cos \left( \tfrac{\phi_s - \phi_i - 2 \phi_p + \phi_1 + \phi_2}{2} \right)
	a_m e^{2 \pi i \fsgnl t} \\
	&-\frac{2 \sqrt{G_s}  e^{i (\phi_s + \phi_i + 2 \phi_p + \phi_2 - \phi_1) / 2}}
	{1 - 2 i (\fsgnl - \fpmp) / \bw}
	\cos \left( \tfrac{\phi_s - \phi_i - 2 \phi_p + \phi_1 + \phi_2}{2} \right)
	a_m e^{2 \pi i (2 \fpmp - \fsgnl) t},
\end{aligned}
\end{equation}
\end{widetext}
which makes the dependence of the gain on the phase of $\phi_1$ and $\phi_2$ evident.
This phase-sensitivity may be advantageous or disadvantageous, depending on the application, such as squeezing.
Phase-insensitive operation can always be achieved by operating with signals on one side of the pump only, i.e. contained only in either the upper or lower amplification side band.

\section{Illustrative example}\label{sec:illustrative_example}

\subsection{Model definition}\label{sec:sm_model_definition}

\begin{figure}
\centering
\includegraphics{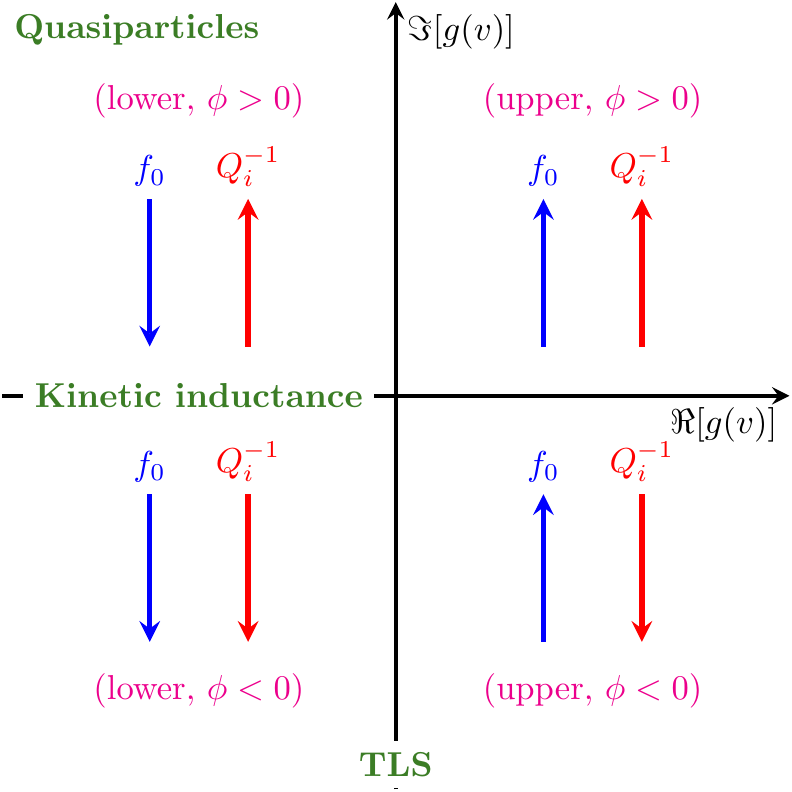}
\caption{\label{fig:quadrant_plot}
Different behaviours modelled by (\ref{eqn:sm_g}), showing the possible locations of $g(v)$ in the Argand plane.
The choice of upper or lower sign in (\ref{eqn:sm_g}), together with the value of $\phi$, determines which quadrant of the plane $g(v)$ will occupy; these combinations are shown in the sets of brackets.
Each quadrant contains a pair of arrows that indicate the corresponding behaviour of the resonant frequency (blue arrow) and internal quality factor (red arrow) with increasing readout power.
If the arrow points upwards the quantity increases with readout power and the opposite when it points downwards.
The green text indicates the location of different physical processes in the model space.
}
\end{figure}

We will now illustrate the formalism described in the preceding sections with a specific simple model.
Assume
\begin{equation}\label{eqn:sm_g}
	g(v) = \pm \frac{v}{v_*} e^{\pm i \phi}
\end{equation}
and
\begin{equation}\label{eqn:sm_h}
	h(v, U) = -\frac{v}{\tau} + \frac{v_*}{\tau} \frac{U}{U_*}
\end{equation}
where $|\phi| < \pi / 2$ and $\tau$, $U_*$ and $v_*$ are real constants.
(\ref{eqn:v_state_equation}) has a straightforward solution for (\ref{eqn:sm_h}), for which (\ref{eqn:model_non_linearity}) yields
\begin{equation}\label{eqn:generalised_duffing}
	\frac{\delta\fres}{\fres} + \frac{i \delta Q_i^{-1}}{2}
	= \pm \frac{e^{\pm i \phi}}{\tau U_* Q_r}
	\int_0^t U(t') e^{-(t - t') / \tau} \, dt'.
\end{equation}
From (\ref{eqn:generalised_duffing}), we see that the model nonlinearity described by (\ref{eqn:sm_g}) and (\ref{eqn:sm_h}) can be viewed as a generalisation of the Duffing nonlinearity\,\cite{swenson2013operation} that allows for rate-limited response and both reactive and dissipative effects.
$U_*$ sets the energy scale of the nonlinearity, $\tau$ is the response time and $\phi$ controls the ratio of reactive to dissipative response.
The assumed sign in (\ref{eqn:sm_g}) sets the direction of the reactive response; for positive sign the resonant frequency frequency increases with stored energy, while for negative sign it decreases.
The sign of the prefactor and exponent in (\ref{eqn:sm_g}) must be chosen consistently; they must either both be positive, or both be negative.
The nonlinearity can be made purely reactive or dissipative by choosing $\phi=0$ or $\phi = \pm \pi/2$ respectively.
For $\phi > 0$, $Q_i$ decreases ($Q_i^{-1}$ increases) with stored energy for both sign choices, while for $\phi < 0$ it increases.
These different possibilities are summarised in Figure \ref{fig:quadrant_plot}.

(\ref{eqn:generalised_duffing}) provides a good approximation to many parametric processes in superconducting resonators, which will necessarily be linear in the stored energy, $U$, to first order.
For example, the kinetic inductance behaviour described by Swenson\,\cite{swenson2013operation} corresponds to  $\phi=0$, $\tau \rightarrow 0$ and assumed negative sign.
Similarly, processes dependent on quasiparticle generation, such as device heating and non-equilibrium behaviour\,\cite{goldie2012non}, are modelled by the case of negative sign, $\tau$ finite and $\phi > 0$, as excess quasiparticles take time to decay, reduce the kinetic inductance and increase the resistance\,\cite{zmuidzinas2012superconducting}.
Finally, TLS effects are a rare example of a process where $Q_i$ increases with stored energy and there is minimal reactive response, so $\phi \approx -\pi / 2$\,\cite{gao2008physics}.
Care should be taken when applying the model in the case $\phi < 0$, as $Q_i$ is unbounded below and can potentially become negative.
However, this simply reflects the fact it is a low-order approximation with a domain of validity; in a full-model of a particular process higher order terms would act to limit $Q_i$.
The green labels in Figure \ref{fig:quadrant_plot} illustrate the general location of $g(v)$ in the Argand plane for each of the physical processes described.

In what follows we will work in terms of $\Gamma_p$ and $\Gamma_s$, as defined earlier, since these are general to any device.
$\Gamma_p$ and $\Gamma_s$ can  be related to the scattering parameters measured for the full device using (\ref{eqn:sp}) and (\ref{eqn:signal_scattering_matrix}).
As discussed in Section \ref{sec:ss_scalar_eqn}, we can picture the full device as a single-port nonlinear resonator in a multi-port embedding circuit.
When pumped appropriately, the underlying single-port resonator functions as a reflection amplifier with gain characterised by the signal frequency reflection coefficient $\Gamma_s$.
Further, we will normally plot quantities derived from $(1 - \Gamma_n)$ rather than from the reflection coefficient $\Gamma_n$ itself.
The reasons for this are two-fold.
First, $(1 - \Gamma_n)$ is typically more directly experimentally accessible, being proportional to the difference between the measured scattering parameters and those expected for the readout circuit, e.g. (\ref{eqn:sp}).
Second, $|1 - \Gamma_n|$ often has more obivious resonant structure then $|\Gamma_n|$.
The most extreme example of the latter point is when the device is lossless and the nonlinearity is reactive and so $|\Gamma_p|=1$ for all frequencies and powers.

\subsection{Operating point}\label{sec:im_op_point}

The possible operating points under pumping are found by working through the analysis of Section \ref{sec:response_to_pump} for $h(v, U)$ and $g(v)$ as given by (\ref{eqn:sm_h}) and (\ref{eqn:sm_g}).
It follows straightforwardly from (\ref{eqn:sm_h}) that the steady state operating condition,
(\ref{eqn:op_point_equation}), requires
\begin{equation}\label{eqn:v_in_ss}
	\frac{v_0}{v_*} = \frac{U_0}{U_*}.
\end{equation}
Using (\ref{eqn:def_p}), (\ref{eqn:sm_g}) and (\ref{eqn:v_in_ss}), we then find
\begin{equation}\label{eqn:ill_mod_p}
	p = 1 + 2 i \left\{ y_p \mp \frac{U_0}{U_*}  e^{\pm i \phi} \right\}.
\end{equation}
The operating points of the device can then be found as the self consistent solutions of (\ref{eqn:ill_mod_p}) and (\ref{eqn:U0_norm}), where we now assume $U_*$ in the definition of $a_*$, (\ref{eqn:def_a_star}), is the scale energy in the model.
Similarly, it follows from (\ref{eqn:def_alpha}), (\ref{eqn:def_beta}) and (\ref{eqn:def_q}) in sequence that
\begin{equation}\label{eqn:ill_mod_alpha}
	\alpha = 1,
\end{equation}
\begin{equation}\label{eqn:ill_mod_beta}
	\beta = \pm \frac{U_0}{U_*} e^{\pm i \phi}
\end{equation}
and
\begin{equation}\label{eqn:ill_mod_q}
	q = \mp \frac{2 i U_0}{U_*} e^{\pm i \phi}
\end{equation}
for the model.

Although we can solve directly for $U_0$, it will prove more convenient to work with an alternate set of variables.
Let
\begin{equation}\label{eqn:def_z}
	z = (1 + 2 i y_p) e^{\mp i \phi},
\end{equation}
\begin{equation}\label{eqn:def_kp}
	k_p = \frac{\Im[z]}{2 \Re[z]}
\end{equation}
and
\begin{equation}\label{eqn:def_k}
	k = k_p \mp \frac{1}{\Re[z]} \frac{U_0}{U_*}.
\end{equation}
The problem of finding the operating point now becomes one of determining $k$ for known $k_p$ (as determined by $y_p$ and $\phi$) and known readout power.
Using (\ref{eqn:ill_mod_p}) and (\ref{eqn:ill_mod_q}), $p$ and $q$ can be expressed in terms of the new variables as
\begin{equation}\label{eqn:sm_p}
	p = \{ 1 + 2 i k \} \Re[z] e^{\pm i \phi}
\end{equation}
and
\begin{equation}\label{eqn:sm_q}
	q= 2 i \{ k - k_p \} \Re[z] e^{\pm i \phi}.
\end{equation}
Using (\ref{eqn:sm_p}) to substitute for $p$ in (\ref{eqn:U0_norm}) and then rearranging, we find $k$ and $k_p$ at the operating point must satisfy
\begin{equation}\label{eqn:sm_state_equation}
	k_p = k + \frac{a}{1 + 4 k^2},
\end{equation}
where
\begin{equation}\label{eqn:def_a}
	a = \pm \frac{a_*}{\Re[z]^3}.
\end{equation}
The main benefit of this transformation is that it puts the operating point equation for the mixed reactive-dissipative nonlinearity into the same form as that for a purely reactive nonlinearity.
The latter has been discussed in the literature\,\cite{swenson2013operation,thomas2020nonlinear}, and many of these results will carry over to the present analysis.
$k_p$ and $k$ can be thought of, respectively, as the `applied' detuning of the pump, as set by $y_p$, and the `realised' detuning, which is the value of the detuning that would give $\Gamma_p$.
This analogy is exact in the limit $\phi = 0$ and follows the definitions used by Swenson\,\cite{swenson2013operation}.
The parameter $a$ controls the degree of nonlinearity, with linear behaviour in the limit $a \rightarrow 0$.

\begin{figure}
\centering
\includegraphics{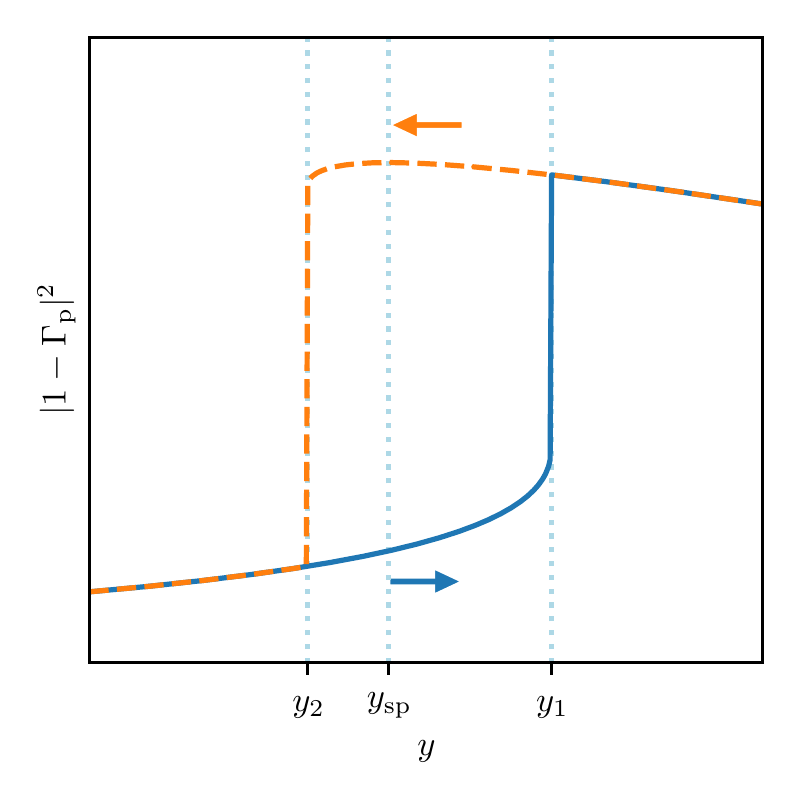}
\caption{\label{fig:generic_switching_curve} Generic form of $|1 - \Gamma_p|^2$ as a function of $y_p$ in the hystereric regime and assuming negative sign in (\ref{eqn:sm_g}).
The blue (solid)  and orange (dashed) curves shows the behaviour that would be measured if $y$ were swept in the the positive and negative $y$-directions respectively.
$y_1$, $y_\text{sp}$ and $y_2$ are all less than zero in this case.
}
\end{figure}

(\ref{eqn:sm_state_equation}) results in either one or two stable operating states for given $y_p$ and $a_*$.
In the case of two stable states the device will display hysteresis, i.e. it is possible to measure different values of $\Gamma_p$ depending on whether the resonant frequency is swept upwards or downwards.
Figure \ref{fig:generic_switching_curve} illustrates the generic form of $|1 - \Gamma_p|^2$ as a function of $y_p$ in the hysteretic regime, assuming negative signs in (\ref{eqn:sm_g}).
On sweeping $y_p$ up in frequency, $|1-\Gamma_p|^2$ follows the lower, solid, curve.
This curve is defined by the possible values of $k$ for which $|k-k_p|$ is minimized.
On sweeping $y_p$ up in frequency, $|1-\Gamma_p|^2$ follows the upper, dashed, curve.
This curve is defined by the possible values of $k$ for which $|k-k_p|$ is maximized.
The two switching points $y_p = y_1$ and $y_p = y_2$ are labelled, along with the stationary point at $y_p = y_\text{sp}$.
If positive sign is assumed in (\ref{eqn:sm_g}), the curve simply transforms according to $y_p \rightarrow -y_p$.

\begin{figure*}
\centering
\includegraphics[scale=0.965]{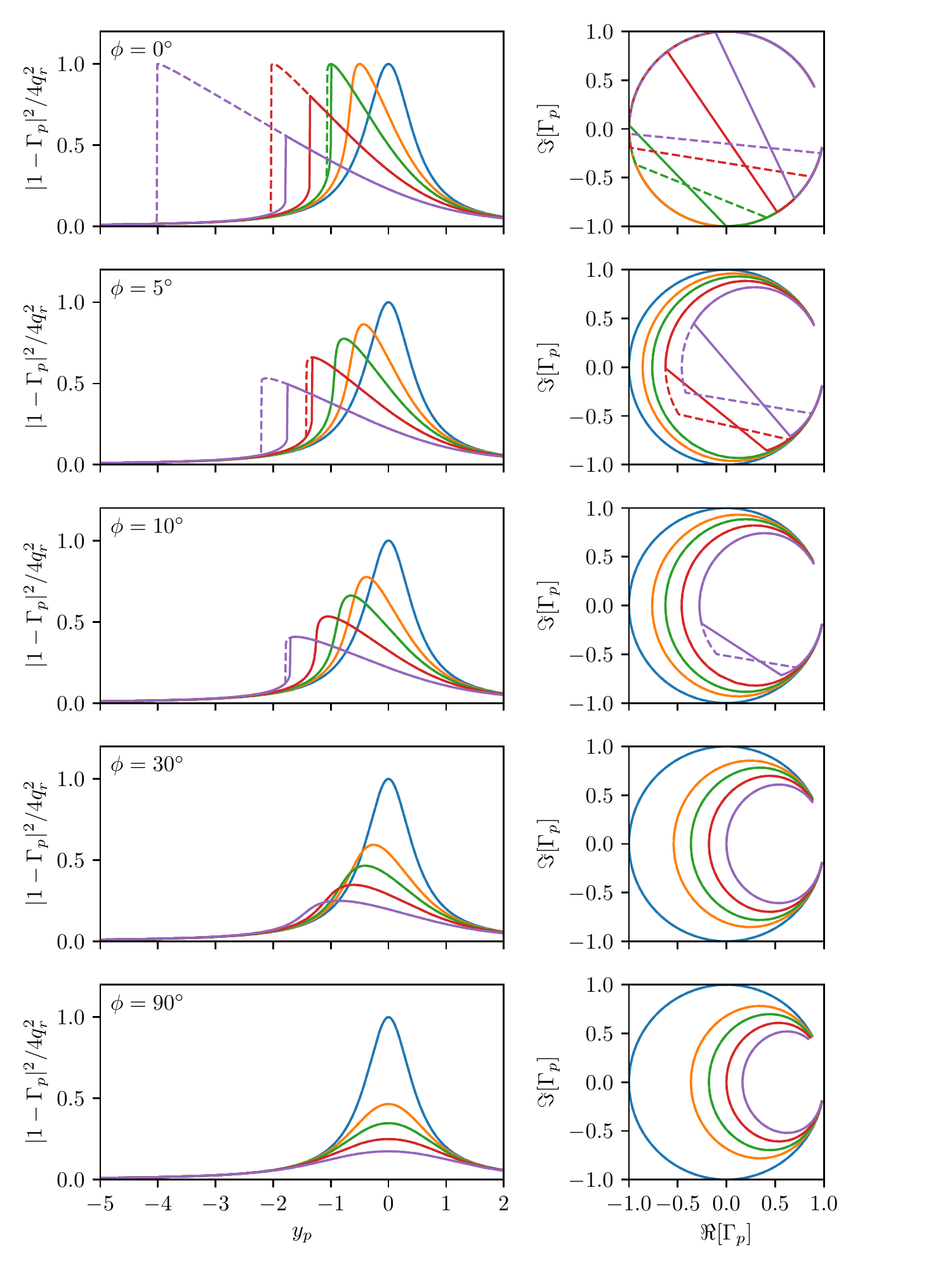}
\caption{\label{fig:pump_curves_different_phi}
Behaviour of $\Gamma_p$ for different values of $a_*$ and $\phi$.
The solid lines are obtained if $y_p$ is swept upwards and the dashed lines if it is swept downwards.
The mappings of line colour to value of $a_*$ are as follows: $\text{blue}=0$, $\text{orange}=0.5$, $\text{green}=1$, $\text{red}=2$ and $\text{purple}=4$.
}
\end{figure*}

Figure \ref{fig:pump_curves_different_phi} illustrates the pumped behaviour of the model resonator over a wide parameter space, assuming negative sign in (\ref{eqn:sm_g}).
It is intended to reflect the data that would be collected in the type of experimental measurements described at the end of Section \ref{sec:response_to_pump} for a typical superconducting resonator.
Each of the rows of the figure corresponds to a different assumed value of the nonlinearity angle $\phi$, while the different line colours within each row correspond to different values of the normalised readout power $a_*$.
The curves were generated by solving (\ref{eqn:sm_state_equation}) for the allowed values of $k$ for given $y_p$ and $a_*$, then using (\ref{eqn:sm_p}) and (\ref{eqn:gamma_p_norm}) to calculate $\Gamma_p$.
Where there are multiple possible values of $k$, the solid/dashed lines of the same colour show the minimum/maximum value of $k$ for a given pump power as accessed by sweeping the pump frequency up/down.
Within each row, the left-hand panel shows $|1 - \Gamma_p|^2$ as a function of $y_p$ and the right-hand panel $\Gamma_p$ over the same range of values of $y_p$ as plotted in the Argand plane.
In the case of a purely linear device, $\Gamma_p$ is expected to follow a circular path in the Argand plane.

Moving down the rows of Figure \ref{fig:pump_curves_different_phi} we see how the behaviour of the resonator changes as the character of the nonlinearity changes from purely reactive ($\phi = 0^\circ$) to purely dissipative ($\phi=90^\circ$).
In the purely reactive case (top row), the model simply reproduces Swenson's results and significant hysteresis is observed.
As the the nonlinearity angle increases we then see mixed behaviour, with shifts in resonant frequency accompanied by increasing dissipation, as indicated by decreasing $|1 - \Gamma_p|^2$.
The presence of any degree of dissipative nonlinearity is seen to produce a characteristic effect in the Argand plane, whereby the resonance `circle' becomes compressed into an oval at sufficiently high readout power.
For $\phi \geq 30^\circ$ we also then see the cessation of hysteresis and we will discuss this behaviour in more detail in the next section.

\subsection{Switching-points}\label{sec:ill_mod_switching_points}

The switching points of the pumped resonator, as introduced in the previous section, play a critical role in amplification.
By graphical argument (e.g. \cite{swenson2013operation, thomas2020nonlinear}) we can show that they occur where $dk_p/dk$ is zero.
Differentiating (\ref{eqn:sm_state_equation}) with respect to $k$ and rewriting $a$ in terms of $k$ and $k_p$, we find
\begin{equation}\label{eqn:deriv_kp_wrt_k}
	\frac{dk_p}{dk}
	= \frac{12 k^2  - 8 k_p k + 1}{1 + 4k^2}.
\end{equation}
Hence the values of $k$ at the switching points must be solutions of the equation
\begin{equation}\label{eqn:sm_sp_eqn}
	12 k^2 - 8 k_p k + 1 = 0.
\end{equation}
However, after some straightforward algebra it follows from (\ref{eqn:sm_p}) and (\ref{eqn:sm_q}) that
\begin{equation}\label{eqn:p_q_rel_at_sp}
	|p + q|^2 - |q|^2 = \{ 12 k^2 - 8 k_p k + 1 \} \Re[z]^2.
\end{equation}
This means $|p + q|^2 - |q|^2 =0$ at the switching points, so by the arguments of Section \ref{sec:amplification_zero_r} they must also be points of infinite small signal gain.
Physically we would expect this, as at the switching points $\Gamma_p$ changes discontinuously for even very small changes in power or pump frequency.
Crucially, this means the parameter space near the switching points is likely to be a good region in which to operate the device as an amplifier.

Unfortunately, it is relatively complicated to calculate the values of $y_p$ at which the switching points will occur for given readout power.
However, the inverse problem, that of determining the readout power necessary to make the device switch at a given frequency, has a straightforward solution.
Assume we want there to be a switching point at a given value of $y_p$ and $\phi$ is known, so that $k_p$ is known.
We can solve (\ref{eqn:sm_sp_eqn}) for $k$ to give
\begin{equation}\label{eqn:sp_k_vals}
	k = \frac{k_p}{3} \pm \frac{1}{3} \sqrt{ k_p^2 -\frac{3}{4}},
\end{equation}
then calculate the necessary power using
\begin{equation}\label{eqn:a_from_k_and_k_p}
	a_* = \pm (k_p - k) (1 + 4 k^2) \Re[z]^3,
\end{equation}
where the latter result follows by rearranging (\ref{eqn:sm_state_equation}).
Note that the $\pm$ in (\ref{eqn:sp_k_vals}) corresponds to the two possible solutions, rather than the choice of sign in (\ref{eqn:sm_g}).
When there are two possible values of $a_*$, the largest and smallest of the values correspond to operating at $y_1$ and $y_2$ in Figure \ref{fig:generic_switching_curve}, respectively.

At the value of $a_*$ above which the resonator response becomes hysteretic, there will exist a single value of $y_p$ for which the two solutions given by (\ref{eqn:sp_k_vals}) are the same --- we will call this the critical point.
At the critical point we must have $|k_p| = \sqrt{3} / 2$ and $k = k_p / 3$.
The condition $a_* \geq 0$ adds the additional requirement $\Im [z] > 0$ for the choice of positive sign in (\ref{eqn:sm_g}) and $\Im[z] < 0$ for negative sign; both choices result in the same final outcome in this analysis, so we will assume the positive sign going forward for convenience.
Taken together, the different requirements constrain the argument of $z$ to be either $\pi / 3$ or $2 \pi / 3$.
When $\phi \geq 0$ then
\begin{equation}
	\text{Arg}[z] = \tan^{-1} (2 y_p) - \phi < \frac{\pi}{2} - \phi,
\end{equation}
so $\text{Arg}[z]$ must be $\pi / 3$ and the critical point can only be accessed if $\phi \leq \pi / 6$.
It must also be the case that
\begin{equation}\label{eqn:yp_critical_point}
	y_p = \pm \frac{1}{2} \tan \left( \frac{\pi}{3} + \phi \right),
\end{equation}
and that the threshold power for bifurcation is
\begin{equation}\label{eqn:threshold_power}
	a_c = \frac{4}{3 \sqrt{3}} |\Re[z]|^3
	= \frac{1}{ 6 \sqrt{3} \cos^3 \left(\frac{\pi}{3} + \phi \right)}.
\end{equation}
A critical point of preceding analysis is that it is not possible to drive the resonator into a hysteretic regime when $\phi > \pi / 6$ (30$^\circ$), as seen in the data in Figure \ref{fig:pump_curves_different_phi}.

This analysis shows that gain is only possible if the amount of dissipative nonlinearity is smaller than the amount of reactive nonlinearity such that $\phi < \pi / 6$ (30$^\circ$).
This threshold is seen in all of our simulations, and is a valuable measure of when gain is possible.
It should also be noted that if two nonlinear mechanisms are present simultaneously, then because the complex frequencies are additive, the overall process is characterised by the vector sum of the two perturbations to the operating point, pole of the pumped resonator, in the complex plane. 
Thus it may be that although one process does not allow gain, the inclusion of another process achieves $\phi < 30^\circ$ and allows gain to be achieved.

\subsection{Experimental tests}\label{sec:ill_mod_exp_test}

Before considering amplification, it is instructive to consider ways of testing how well the model describes a real device.
One way to do so is using swept frequency/power measurements of the type simulated in Figure \ref{fig:pump_curves_different_phi}.
The model predicts that $|1 - \Gamma_p|^2$ will be observed to have a stationary point (specifically a maximum) with respect to $y_p$ for given readout power, provided the measurement frequency is swept in the same direction as the resonator moves with increasing readout power.
This is illustrated in Figure \ref{fig:generic_switching_curve} and we let  $y_s$ and $\Gamma_s$ denote the values of $y_\text{sp}$ and $\Gamma_p$ at the stationary point, respectively.
In Section \ref{sec:stationary_points} of the supplementary material we show that at this point
\begin{equation}\label{eqn:ys_stat_point}
	y_\text{sp} = \pm \frac{a_* \cos \phi}{|p|^2}
\end{equation}
where the sign is determined by that in (\ref{eqn:sm_g}).
Using the definitions of $y_p$, $a_*$ and $p$ we can rewrite (\ref{eqn:ys_stat_point}) in the more experimentally useful form
\begin{equation}\label{eqn:fs_stat_point}
	f_\text{sp} = \fres \pm |1 - \Gamma_s|^2 \fres P_0 \cos \phi / P_*,
\end{equation}
where $f_\text{sp}$ is the measurement frequency corresponding to $y_\text{sp}$ and
\begin{equation}
	P_* = \frac{4 Q_r^3 P_0}{Q_c^2 a_*}
	= \frac{2\pi \fres Q_r U_* |\mathbf{k}|^2 |\mathbf{a}_0|^2}
	{Q_c| \mathbf{k}^T \! \! \cdot \mathbf{a}_0|^2}
\end{equation}
is a scale power.
Hence we can test for compliance by taking a series of frequency sweeps at several different values of the pump power $P_0$, plotting $f_\text{sp}$ versus $|1 - \Gamma_s|^2 P_0$ and then checking whether the points lie on a straight line.

If agreement is found, we can go further and determine model parameters.
Values for $\fres$ and $P_* \cos \phi$ can be obtained by be obtained by fitting the straight line plot just described.
Similarly, by fitting very low power data we can also obtain $Q_r$, which can be used with $\fres$ to convert measurement frequencies into values of the detuning $y_p$.
Once we are able to do this conversion, by using the measured frequency location of the critical point and (\ref{eqn:yp_critical_point}) we can then find $\phi$.
This in turn allows $P_*$ to be calculated, along with the scale energy $U_*$ if the coupling quality factor $Q_c$ is known.

\subsection{Amplification}\label{sec:sm_amplification}

\begin{figure*}
\centering
\includegraphics[]{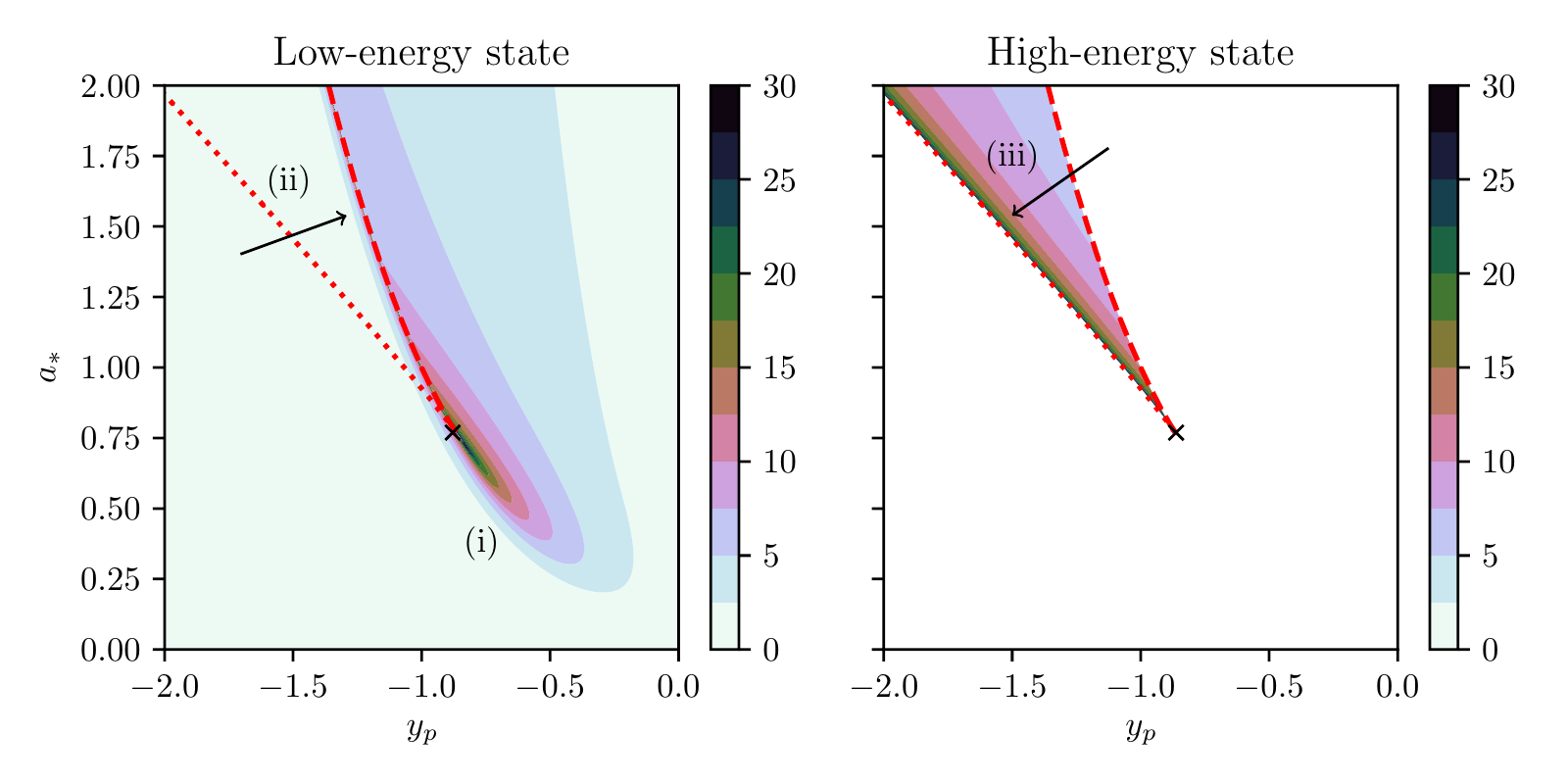}
\caption{\label{fig:gain_contour_plot_phi_zero}
Filled contour plots showing maximum signal gain $|\Gamma_s|^2$ with respect to $y_s$, in dB, plotted as a function of $a_*$ and $y_p$ for $q_r = 1$, $\phi = 0$ and assumed negative sign in (\ref{eqn:sm_g}).
The red dashed lines show the location of the switching points in behaviour with respect to $y_p$.
Behaviour is shown for both the high and low energy states of the pumped resonator.
A separate high-energy state only exists in pump power ranges where the resonator state can bifurcate and in this example is accessed by sweeping $y_p$ down from $y_p > 0$ to the desired value.
The regions where the state does not exist are coloured white in the right-hand plot.
The location of the critical point discussed in the text is indicated by the black diagonal cross in each panel.
}
\end{figure*}

Once the operating points of the states of the resonator have been found, the analysis of Section \ref{sec:small_signal_analysis} can be used to calculate the behaviour as an amplifier.
In practice we have control over the applied pump power $a_*$ and the pump detuning $y_p$.
In this section we will consider the achievable amplification as a function of these two variables, which will highlight operating regions for further discussion.

Consider the case of a purely reactive nonlinearity ($\phi = 0$) and negative assumed sign in (\ref{eqn:sm_g}).
This is the behaviour that would be expected of a superconducting resonator in which the intrinsic kinetic inductance nonlinearity is dominant.
For each choice of $a_*$ and $y_p$, there will be some value of the signal detuning $y_s$ for which the small signal gain as given by $|\Gamma_s|^2$ is maximised.
Figure \ref{fig:gain_contour_plot_phi_zero} shows the calculated value of this maximum gain, in decibels (dB), as a function of $a_*$ and $y_p$.
For values of $a_*$ and $y_p$ where the resonator has two stable operating states the gain can take two different values.
The left-hand panel in Figure \ref{fig:gain_contour_plot_phi_zero} shows the behaviour expected when the resonator is in either the single available operating state or, where two states exist, the state corresponding to the smallest stored energy.
This is the state corresponding to the smallest value of $|k|$, i.e. the device follows the blue line in Figure \ref{fig:generic_switching_curve}.
The two broken red lines show the location of the switching points with respect to the pump and bound the region with multiple operating states, which we will call the bifurcated region.
The low energy state is accessed by approaching the bifurcation region in the direction of the arrow, i.e. from lower $y_p$ or $a_*$.
The right-hand panel in Figure \ref{fig:gain_contour_plot_phi_zero} shows the gain in the high-energy state where the state exists.
This has the higher value of $|k|$ of the two states and corresponds to the resonator following the orange curve in Figure \ref{fig:generic_switching_curve}.
It is accessed by entering the bifurcation region from higher $y_p$ or $a_*$, as indicated by the arrow.
The colour scale in both panels has been cut off at 30\,dB for clarity.
Small regions with gain in excess of  30\,dB  exist in both plots, but much of the detail of the larger scale structure is lost if the colour scale  is extended to accommodate them.

Figure \ref{fig:gain_contour_plot_phi_zero} indicates three regions of high gain.
The first, labelled (i), is found near the critical point where the resonator behaviour first bifurcates with increasing power.
The second, labelled (ii), is in the low-energy state of the bifurcated region where the pump is approaching the relevant switching point indicated by the dashed red line.
Although not immediately obvious from the plot, we will show  shortly that the gain becomes infinitely large as the red-dashed line is approached.
The third region, labelled (iii), is in the high-energy state in the bifurcated region.
Gain is uniformly high in this region, but tends to very large values as the switching point indicated by the red dotted line is approached.

\begin{figure*}
\centering
\includegraphics[]{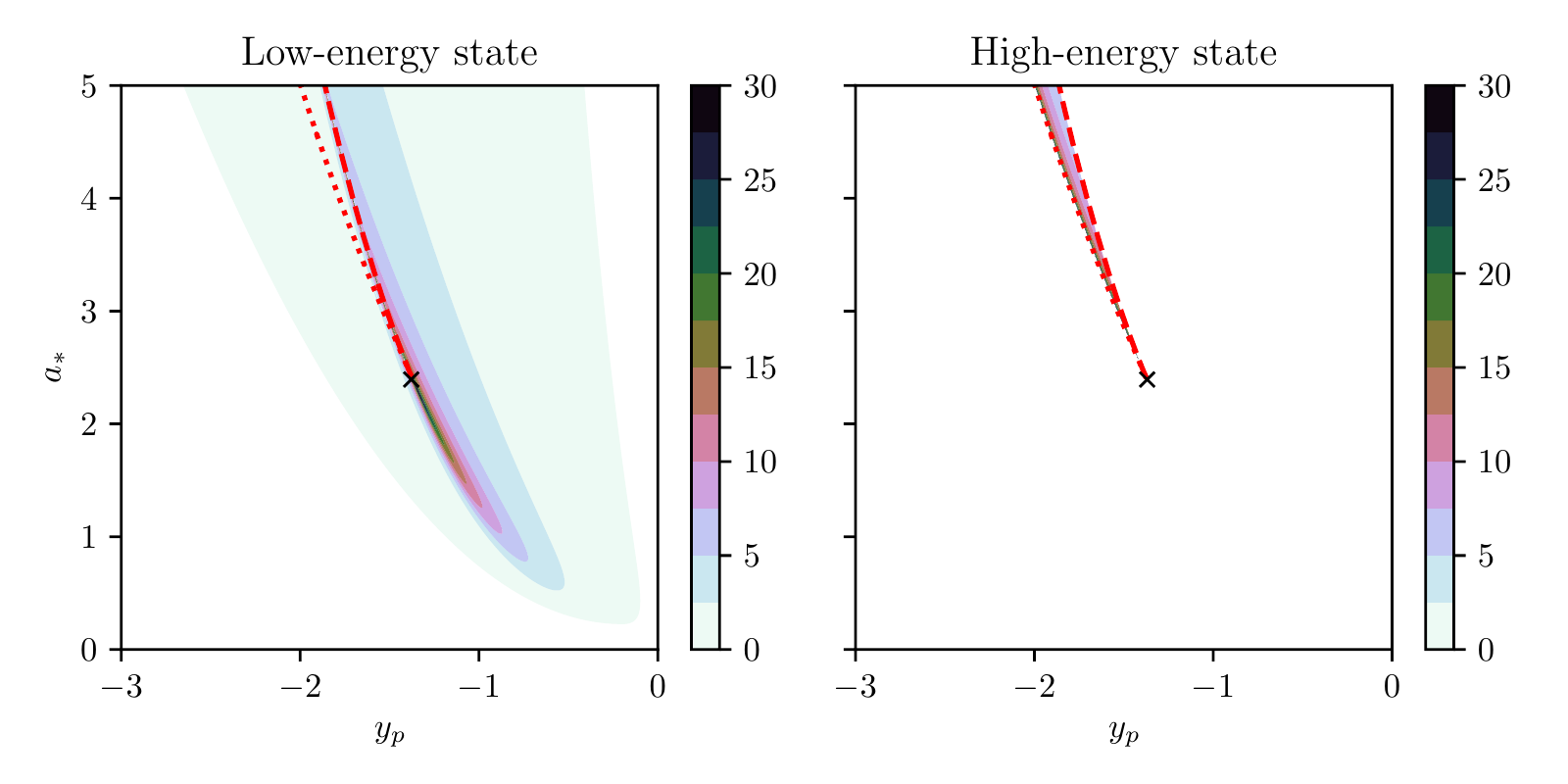}
\caption{\label{fig:gain_contour_plot_phi_10deg}
Figure \ref{fig:gain_contour_plot_phi_zero} for $\phi=10^\circ$ and wider range of $y_p$.
The white areas in the plot for the low-energy state indicate regions where the maximum gain is below 0\,dB.
}
\end{figure*}

Figure \ref{fig:gain_contour_plot_phi_10deg} shows the equivalent plot to Figure \ref{fig:gain_contour_plot_phi_zero} for the case where $\phi =10^\circ$.
This is more representative of amplification processes involving quasiparticles in a superconducting resonator, since any change in resonant frequency must be accompanied by a decrease in internal quality factor.
For positive $\phi$ is possible for $|\Gamma_s|^2$ to fall below 0\,dB and so as to maintain reasonable colour scales we have coloured these regions white in the left-hand panel of the figure.
The behaviour for $\phi =10^\circ$ is broadly similar to that for $\phi =0^\circ$ in the sense that regions of reasonable gain, $>10\,\text{dB}$, still exist.
However, critically these regions are seen to be much smaller for $\phi =10^\circ$ than when $\phi =0^\circ$ and for higher powers to be needed to access them.
The former implies fine tuning of the operating parameters and potentially stability issues, while the latter increases the risk of exciting other nonlinear processes.
In particular, the region over which the favourable high-gain, high-energy, state is accessible is greatly reduced.

\subsection{Amplification near the switching points}\label{sec:sp_amplification}

As discussed, highest small-signal gain is expected in the vicinity of the switching points in the behaviour of the pump reflection coefficient.
Figure \ref{fig:generic_switching_curve} therefore suggests two different ways of biasing the device into an amplifying state given appropriate pump power $a_*$.
The first is to sweep $y_p$ up from a large negative value to just below the switching point at $y_1$; alternatively, $y_p$ could be set at the desired value and $a_*$ increased from zero until gain is observed.
The second is to sweep $y_p$ down from a large positive value to just above the switching point a $y_2$.
Different behaviour is expected at the two points, due the differing amounts of energy stored in the resonator.

\begin{figure*}
\centering
\includegraphics[]{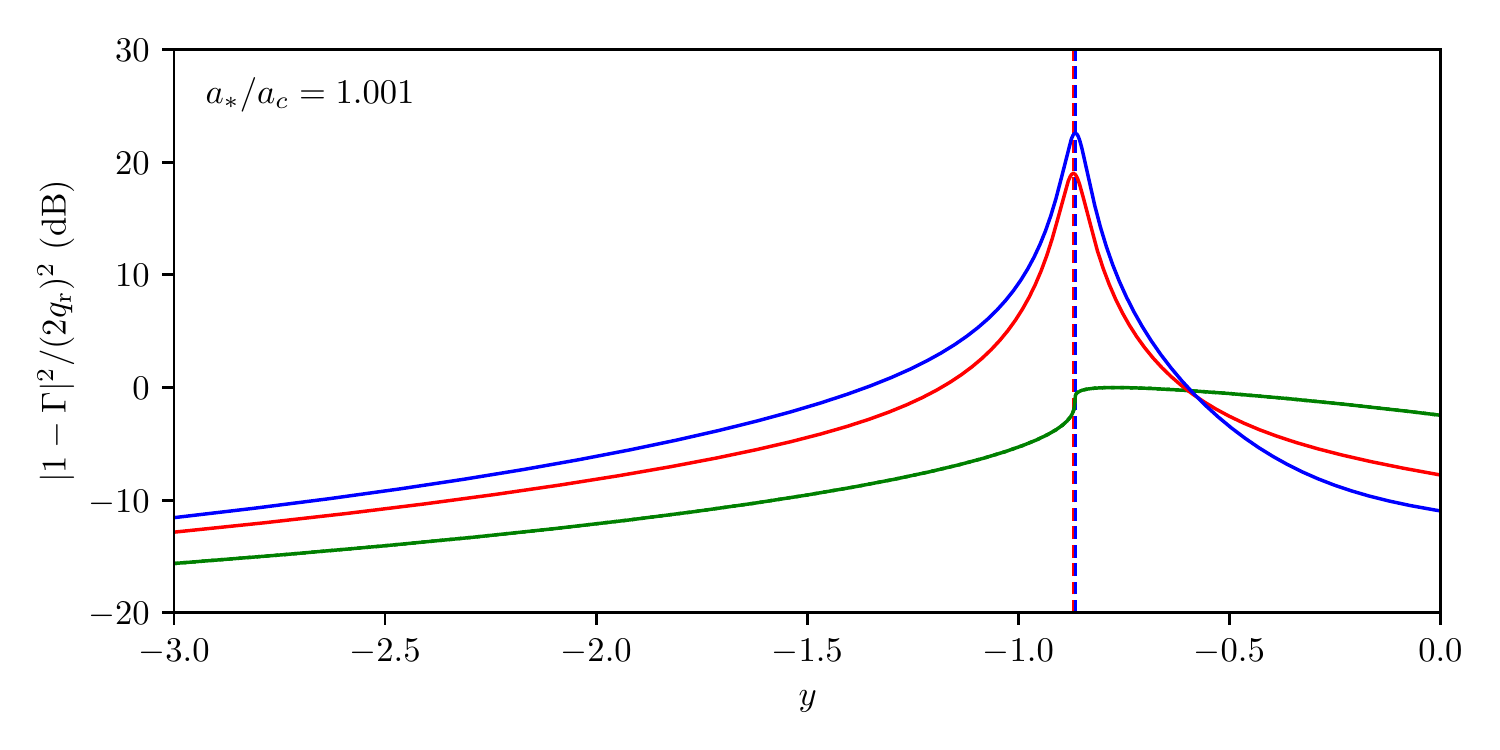}
\caption{\label{fig:amp_plot_1}
Amplification performance for different pump detunings, assuming $\phi=0$, $Q_r / Q_c = 1$ and negative sign in (\ref{eqn:sm_g}).
The green curves shows $\Gamma = \Gamma_p$ for $y_p = y$ when $y$ is swept upwards (solid curve) and downwards (dashed curve), so as to highlight the expected location of the switching points.
The applied readout power has been set to just above the critical power $a_c = 4 / 3 \sqrt{3}$ at which switching is first expected, so $y_1$ and $y_2$ as defined in Figure \ref{fig:generic_switching_curve} are almost identical.
The blue curve shows $\Gamma = \Gamma_s$ for $y_s = y$ with the pump frequency tuned to $2.5 \times 10^{-3}$ above the lower switching point, $y_2$, assuming the pump has been swept down from $y=\infty$; essentially is shows the signal frequency gain as a function of $y$.
Pump location is indicated by the vertical dashed blue line.
The red curve is the equivalent plot with the pump frequency having been tuned to $2.5 \times 10^{
-3}$ below the upper switching point, assuming the pump has been swept upwards from $y=-\infty$.
Pump location for this case is indicated by the vertical dashed red line.
}
\end{figure*}

\begin{figure*}
\includegraphics[]{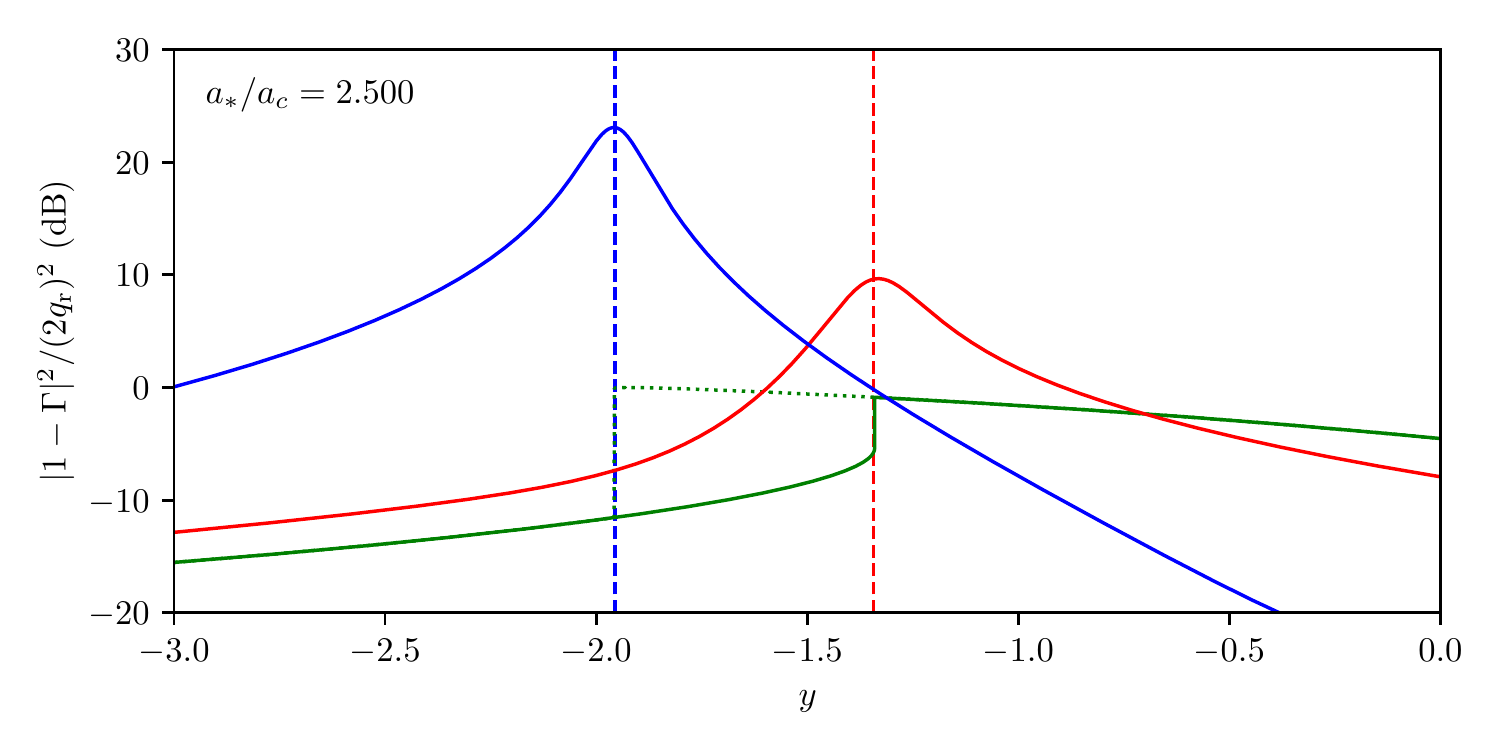}
\caption{\label{fig:amp_plot_2} Equivalent plot to Figure \ref{fig:amp_plot_1}, but with the applied power now increased to $a = 2.5 a_c$.
}
\end{figure*}

Figures \ref{fig:amp_plot_1} and \ref{fig:amp_plot_2} illustrate this behaviour.
In both plots the model has been assumed to infinitely fast and purely reactive; $r$ and $\phi=0$.
In Figure \ref{fig:amp_plot_1} the normalised readout power $a_*$ is just above the threshold for bifurcation, whereas in Figure \ref{fig:amp_plot_2} it is much larger.
$|1 - \Gamma_p|^2 / 2 q_r$ is plotted for $y_p  = y$ for both sweep directions (green solid and dashed lines) to aid with identification of the switching points.
The red and blue lines show the calculated values of $|1 - \Gamma_s|^2 / 2 q_r$ for pump detunings close to the these points.
The exact detunings are shown by the dashed lines of matching colour and have been chosen to be a distance $\delta y = 2.5 \times 10^{-3}$ from the relevant switching point in the appropriate direction.

Applying the results of Section \ref{sec:amplification}, for pump detunings near the switching points we expect to be able to approximate
\begin{equation}\label{eqn:signal_rc}
	\Gamma_s = 1 - \frac{\sqrt{G_0} e^{i \theta_0}}{1 + 2 i (y_s - y_p) / \Delta y_{3\text{dB}}}
\end{equation}
where $G_0$ is the real-valued power gain, $\theta_0 = \text{Arg}[(p + q) / \Re[p + q]]$ and $\Delta y_{3\text{dB}}$ is the 3\,dB bandwidth of $|1-\Gamma_s|^2$ expressed in linewidths.
(\ref{eqn:gain_band_prod}) can be expressed in terms of $\Delta y_{3\text{dB}}$ as
\begin{equation}\label{eqn:sm_gbwp}
	\frac{\sqrt{G_0} \Delta y_{3\text{dB}}}{q_r}
	= \frac{|p + q|}{\Re[p+q]}.
\end{equation}
Further, it can be shown that
\begin{equation}\label{eqn:sm_gain_approx}
	\frac{G_0}{4 q_r^2}
	\approx
	\begin{cases}
		\frac{\sqrt[3]{3}}{36} \frac{1}{|y_p - y_n|^{4/3}} & y_1 = y_2 \\
		\frac{1}{64 y_1 (y_p - y_1)} & y_p \approx y_1 \\
		-\frac{y_2}{4 (y_p - y_2)} & y_p \approx y_2.
	\end{cases}
\end{equation}
and
\begin{equation}\label{eqn:sm_gbwp_approx}
	\frac{\sqrt{G_0} \Delta y_{3\text{dB}}}{q_r}
	\approx \begin{cases}
		\frac{2}{\sqrt{3}} & y_p \approx y_1, y_1 = y_2 \\
		\sqrt{1 + (2 y_1 / 3 )^2} & y_p \approx y_1 \\
		2 |y_2| & y_p \approx y_2
	\end{cases}.
\end{equation}
The derivation of (\ref{eqn:sm_gain_approx}) is purely algebraic; details are given in Section \ref{sec:gain_near_sp} of the supplementary material.
Similarly, the gain-bandwidth products follow by evaluating (\ref{eqn:sm_gbwp}) at the switching points using the approximate relationships between $k_p$, $k$ and $a$ derived in Section \ref{sec:gain_near_sp}.

(\ref{eqn:sm_gain_approx}) shows that the achievable gain is approximately inversely proportional to the distance of the pump from the switching point in all cases.
The case $y_1 = y_2$ in this result corresponds to Figure \ref{fig:amp_plot_1}, i.e. where the device is still near the critical point.
(\ref{eqn:sm_gain_approx}) gives an estimated signal gain of $\approx$24\,dB for the assumed signal detuning, which is very close to that seen in the plot.
Away from the critical point, the switching point at $y_p = y_2$ is seen to offer better performance in terms of scaling of gain with detuning of the pump from the switching point, as shown by (\ref{eqn:sm_gain_approx}), as well as larger gain bandwidth product, as shown by (\ref{eqn:sm_gbwp_approx}).
This is clearly illustrated in Figure \ref{fig:amp_plot_2}, where the blue curve has a much higher peak gain than the red curve.
However, the practical disadvantage of operating at this point is likely reduced stability; when the gain is more sensitive to operating parameters there is greater chance of the signal forcing the device to irreversibly switch states.

\subsection{Bandwidth}\label{sec:bandwidth}

\begin{figure*}
\centering
\includegraphics{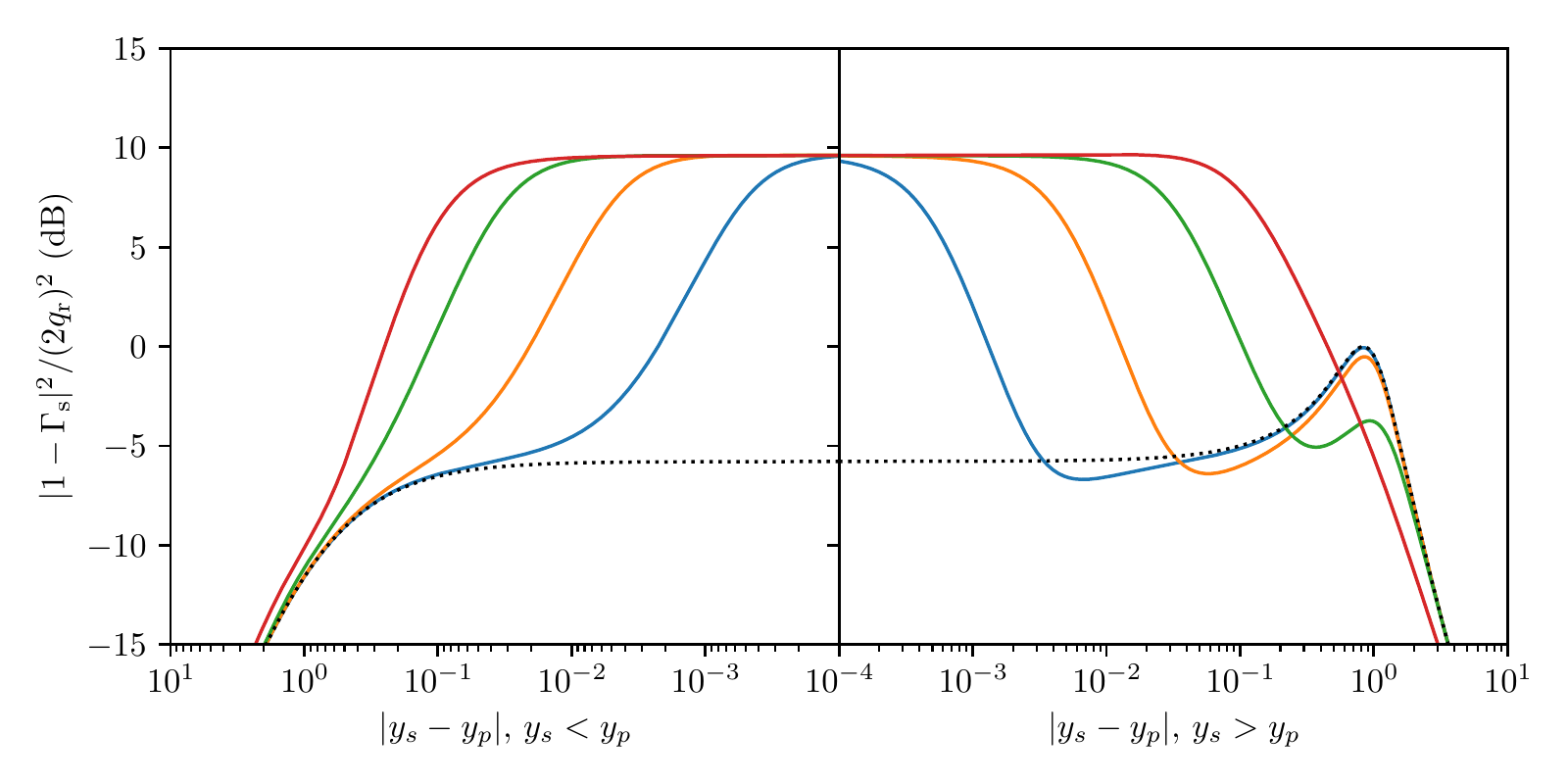}
\caption{\label{fig:bandwidth_plot}
Signal gain around the pump frequency for the same settings as the red curve in Figure \ref{fig:amp_plot_2} and different values of the speed parameter $r$.
The red curve is for $r=0$ (infinitely fast nonlinearity) and is the same as the line in Figure \ref{fig:amp_plot_2}, while the green, orange and blue lines show the results for $r=1$, $10$ and $100$ respectively.
The black dotted line shows the expected response for a linear resonator with resonant frequency and quality factor equal to those of the nonlinear resonator in the pumped state.
}
\end{figure*}

In addition to gain, it is important to consider the spectral response of an amplifier.
The red solid line in Figure \ref{fig:bandwidth_plot} shows the data for the red curve in Figure \ref{fig:amp_plot_2} replotted as a Bode plot with the pump frequency as the zero reference.
In this format the single-pole role-off characteristics predicted by (\ref{eqn:signal_rc}) are clearly visible.
The other lines in the figure illustrate the effect of increasing $r$.
This is physically equivalent to increasing the response time $\tau$ and therefore slowing the parametric process.
Moving from left to right, the value of $r$ increases in factors of ten from $1$ (green) to $10$ (orange) to $100$ (blue).
As predicted in Section \ref{sec:amplification_finite_r}, the maximum gain is unaffected and the bandwidth is seen to be inversely proportional to $r$.

In the limit $r \rightarrow \infty$ we would expect no amplification, but that the signal should still see the behaviour of the resonator as modified by the pump.
The latter case is shown by the black dotted curve in Figure \ref{fig:bandwidth_plot} and we do indeed observe the out of band behaviour tending to the curve as $r$ is increased.
Notice, however, that the amplifier gain always dips below the black dotted line to the right of the band edge above the pump frequency.
This phenomena is characteristic of amplification via a rate-limited nonlinearity and has been discussed in detail by Zhao\,\cite{zhao2022nonlinear}.

\subsection{Amplification by quasiparticle generation}\label{sec:qp_amp_processes}

\begin{figure*}
\centering
\includegraphics[width=\textwidth]{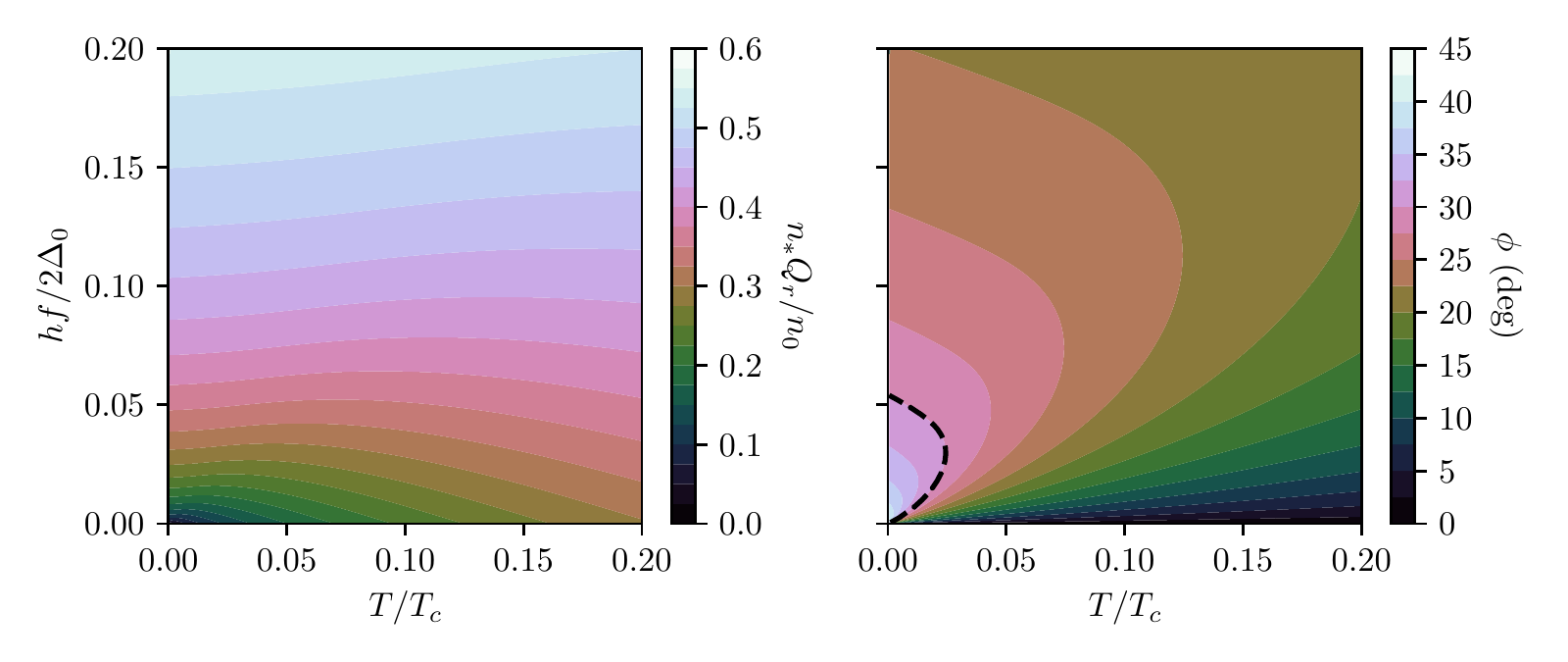}
\caption{\label{fig:qp_params_plot}
Contour plots of $n_*$ (left) and $\phi$ (right) for quasiparticle processes as a function of reduced temperature $T/T_c$ and reduced frequency $h f / 2 \Delta_0$.
The two plots share the same vertical axis.
The dashed black line in the $\phi$-plot indicates the 30$^\circ$ contour.
}
\end{figure*}

If the model described here is applied to superconducting resonators having quasiparticle generation processes, then $\phi$ is non-negative.
The constraint that $\phi < 30^\circ$ to achieve gain, then defines an operating regime for the device.
In this section we will consider the ability to achieve amplification using a superconducting resonator that generates quasiparticles by dynamical pair-breaking.
Although discussions of parametric amplification in superconductors typically focus on the so-called `intrinsic' kinetic inductance nonlinearity in superconductors, we will see that dynamical pair-breaking could be used as an alternate mechanism.

To first-order, the state of the quasiparticles in a superconducting resonator can be represented by a device-averaged quasiparticle density $\nqp$ (see Appendix \ref{sec:hw_resonator} for a specific example).
When a signal is applied to the resonator, a small excess quasiparticle density $\delta \nqp$ over the thermal value results.
Following Zmuidzinas\,\cite{zmuidzinas2012superconducting}, the corresponding shifts in the resonant frequency and Q-factor of the device are
\begin{equation}\label{eqn:df_nqp}
	\frac{\delta \fres}{\fres} = - S_2 \frac{\delta \nqp}{n_0}
\end{equation}
and
\begin{equation}\label{eqn:dq_nqp}
	\frac{\delta Q^{-1}}{2} =  S_1 \frac{\delta \nqp}{n_0},
\end{equation}
where $S_1$ and $S_2$ are dimensionless, and $n_0$ is a scaling density that depends on the material, kinetic inductance fraction of the resonator and the penetration of the resonator fields into the superconductor.
For a near thermal starting distribution, $S_1$ and $S_2$ are approximately
\begin{equation}\label{eqn:def_S1}
	S_1 = \frac{2}{\pi} \sqrt{\rho} \, \sinh (\xi) K_0 (\xi)
\end{equation}
and
\begin{equation}\label{eqn:def_S2}
	S_2 = 1 + \sqrt{\rho} \, e^{-\xi} I_0 (\xi),
\end{equation}
where $\rho = 2 \Delta_0 / (\pi k_b T)$, $\xi = h \fres / 2 k_b T$, $T$ is the temperature of the resonator, $\Delta_0 \approx 1.764 k_b T_c$ is the gap energy of the superconductor at absolute zero, $T_c$ is the superconducting critical temperature, $k_b$ is Boltzmann's constant and $h$ is Planck's constant.
$K_0$ an $I_0$ are the zeroth order modified Bessel functions of the first and second kind, respectively.
(\ref{eqn:df_nqp}) and (\ref{eqn:dq_nqp}) can be conveniently combined into a single expression
\begin{equation}
	\frac{\delta \fres}{\fres} + \frac{i \delta Q^{-1}}{2}
	= -\frac{e^{-i \phi}\delta \nqp}{n_* Q_ r},
\end{equation}
where $\phi = \tan^{-1} (S_1 / S_2)$ and $n_* = n_0 / Q_r \sqrt{S_1^2 + S_2^2}$.
This is of the form of (\ref{eqn:model_non_linearity}), with $\delta \nqp$ as the new state variable.

The contour plots in Figure \ref{fig:qp_params_plot} show $n_*$ and $\phi$ as a function of both reduced temperature $T/T_c$ and reduced frequency $h \nu / (2 \Delta_0)$ over the ranges for which the approximations (\ref{eqn:def_S1}) and (\ref{eqn:def_S2}) are expected to hold.
This range is also reflective of the typical operating point of superconducting microresonators.
The behaviour of $\phi$ is such that amplification can be achieved by quasiparticle generation over a large fraction of the parameter space, with the only exception being the small area at low temperatures and frequencies that is enclosed by the dashed black line.
The normalised scaling density $Q_r n_* / n_0$, which determines necessary pump powers, is seen to vary primarily as a function of frequency.
However, the dependence is relatively weak and the variation is only of order of a factor of three over the majority of the parameter space.
As a result, $n_*$ will be mainly determined by $n_0$, as set by the superconductor, and $Q_r$, as set by a combination of the materials used ($Q_i$) and device geometry ($Q_c$), rather than by the operating point.

Results (\ref{eqn:df_nqp})--(\ref{eqn:def_S2}) follow from considering Owen's\,\cite{owen1972superconducting} chemical potential model for dynamical pair-breaking in the low-temperature, low-frequency, limit\,\cite{gao2008equivalence}.
It is conceptually straightforward, however, to extend the analysis to wider temperature and frequency ranges, actually covering the whole of the superconducting state, but because this requires numerical analysis, we will present this work in a different paper.
Interestingly, all of the plots show that appreciable gain is possible over a very wide range of temperature and frequency, opening up the fascinating possibility of high temperature ($<$10\,K) submillimetre-wave ($<$1\,THz) amplifiers.

\section{Conclusions}\label{sec:conclusions}

We have presented a general formalism for describing the behaviour of resonator parametric amplifiers.
The model allows both reactive and dissipative nonlinear processes to be present simultaneously, and includes the effects of limiting the rate at which the parametric processes can act.
Rather than describing parametric effects in terms of equivalent circuit elements (L,C,R), we describe them in terms of shifts in resonant frequency and reciprocal Q, or equivalently shifts in the pole of the resonator in the complex plane, driven by changes in stored energy.
Within this formalism, amplifier performance is determined by three dimensionless complex parameters: $p$, which characterises the operating point established by the pump; $q$, which characterises the dynamical behaviour around the operating point; and $r$, which characterises the speed of the nonlinearity and determines amplifier bandwidth.
We have shown how these quantities can be calculated from a model, how they can be measured experimentally and how they can be optimized for gain.
Although we have couched our discussions in terms of physical processes in superconducting resonators, the results are generally applicable.

Crucially, to our knowledge, this is the first time the effects of rate-limiting the nonlinearity have been considered.
We have shown that a nonlinear mechanism with finite response time can still produce the same gain as an otherwise equivalent mechanism with instantaneous response.
However, the bandwidth over which it can do so is reduced.
This observation is particularly relevant in superconducting devices, where different physical processes having similar nonlinear behaviour to first order can be distinguished by their different characteristic timescales.

To illustrate the formalism, we have analysed in detail a generalisation of the Duffing model that allows for mixed reactive/dissipative nonlinear behaviour and finite response time.
Although simple, this model and the analytic results obtained are highly valuable for explaining behaviour in real devices.
Indeed, we have shown that there is no obvious reason why quasiparticle generation processes (e.g. readout power heating) cannot be used for amplification in the normal operating regime of most superconducting resonators.
In an upcoming paper we will describe extensive measurements on a set of thin-film amplifiers fabricated from Al, Ti, Nb and NbN.
These devices show behaviour that is entirely consistent with the dynamics revealed here, including the existence of bifurcation points, the procedures needed to bias on the different branches of the hysteresis, the different amplifier characteristics found at the different operating points, and the rate-limiting effects of quasiparticle relaxation.
We are considering many extensions to the model, including the realisation of broad-band amplifiers, based on multi-pole filters.

Finally, one assumption of the model should be restated explicitly.
To this point we have only considered slow nonlinearities, i.e. those that depend on $|u|^2$, as distinct from fast nonlinearities, i.e. those that depend on $u^2$.
This approach was adopted because for sufficiently slow response time, the fast terms could be ignored.
However, in some real devices, both process will be present at some level, and this will affect the realisable gain.
The interplay between these two types of nonlinearity is likely to be a fruitful area of future study.

%apsrev4-2.bst 2019-01-14 (MD) hand-edited version of apsrev4-1.bst
%Control: key (0)
%Control: author (8) initials jnrlst
%Control: editor formatted (1) identically to author
%Control: production of article title (0) allowed
%Control: page (0) single
%Control: year (1) truncated
%Control: production of eprint (0) enabled
%

%%%%% Supplementary material %%%%%

\appendix

\section{Device model}\label{sec:device_model_sm}

\subsection{Proof of the relationship between the readout circuit scattering matrix and the coupling vector}\label{sec:s_and_k_relationship}

The constraint on $\mathbf{k}$ and $\mathsf{S}_0$ given by (\ref{eqn:s_and_k_relationship}) follows from conservation of energy.
Using (\ref{eqn:coupled_output_equation}) we can express the time-averaged power flow into the coupled system as
\begin{equation}\label{eqn:gen_power_in}
	P = |\mathbf{a}|^2 - |\mathbf{b}|^2
	= - |\mathbf{k}|^2 |u|^2 - 2 \Re[ (\mathbf{a}^\dagger \cdot \mathsf{S}_0^\dagger \cdot \mathbf{k}) u],
\end{equation}
where the time dependence of $\mathbf{a}$, $\mathbf{b}$ and $u$ has been suppressed for notational convenience and we have used the fact $\mathsf{S}_0$ is unitary to eliminate terms.
Similarly, we can use (\ref{eqn:coupled_state_equation}) to express the rate of change in the total energy stored in the resonator as
\begin{equation}\label{eqn:gen_energy_deriv}
	\frac{dU}{dt} = 2 \Re \left[ u^* \frac{du}{dt} \right]
	= -\frac{2 \pi \fres}{Q_i} |u|^2 - |\mathbf{k}|^2 |u|^2
	+ 2 \Re [ (\mathbf{a}^\dagger \cdot \mathbf{k}^*) u ],
\end{equation}
where we have used $Q_r^{-1} = Q_i^{-1} + Q_c^{-1}$ to separate the loss term into contributions from coupling- and internal-losses.
Conservation of energy implies that the difference between the net power flow into the system and the rate of change in the stored energy should equal the power dissipated internal to the resonator, as given by $2 \pi \fres |u|^2 / Q_i$.
However, subtracting (\ref{eqn:gen_energy_deriv}) from (\ref{eqn:gen_power_in}) we see that this requires
\begin{equation}\label{eqn:diff_constraint_1}
	\Re[ \mathbf{a}^\dagger \cdot (\mathsf{S}_0^\dagger \cdot \mathbf{k} + \mathbf{k}^*) u] = \mathbf{0}
\end{equation}
for all $u$ and $\mathbf{a}$.
By considering the complementary cases $u \mathbf{a} = \mathbf{v}$ and $u \mathbf{a} = i\mathbf{v}$ for real $\mathbf{v}$, we can show (\ref{eqn:diff_constraint_1}) is equivalent to requiring $\mathbf{v} \cdot (\mathsf{S}_0 \cdot \mathbf{k} + \mathbf{k}^*) = \mathbf{0}$ for all real $\mathbf{v}$.
For non-zero $\mathbf{k}$, we must therefore have
\begin{equation}\label{eqn:diff_constraint_2}
	\mathbf{k}^* = -\mathsf{S}_0^\dagger \cdot \mathbf{k}.
\end{equation}
(\ref{eqn:s_and_k_relationship}) then follows via the unitarity of $\mathsf{S}_0$.

\subsection{State-representation of a capacitively coupled transmission line resonator}\label{sec:hw_resonator}

\begin{figure*}
\centering
\includegraphics[]{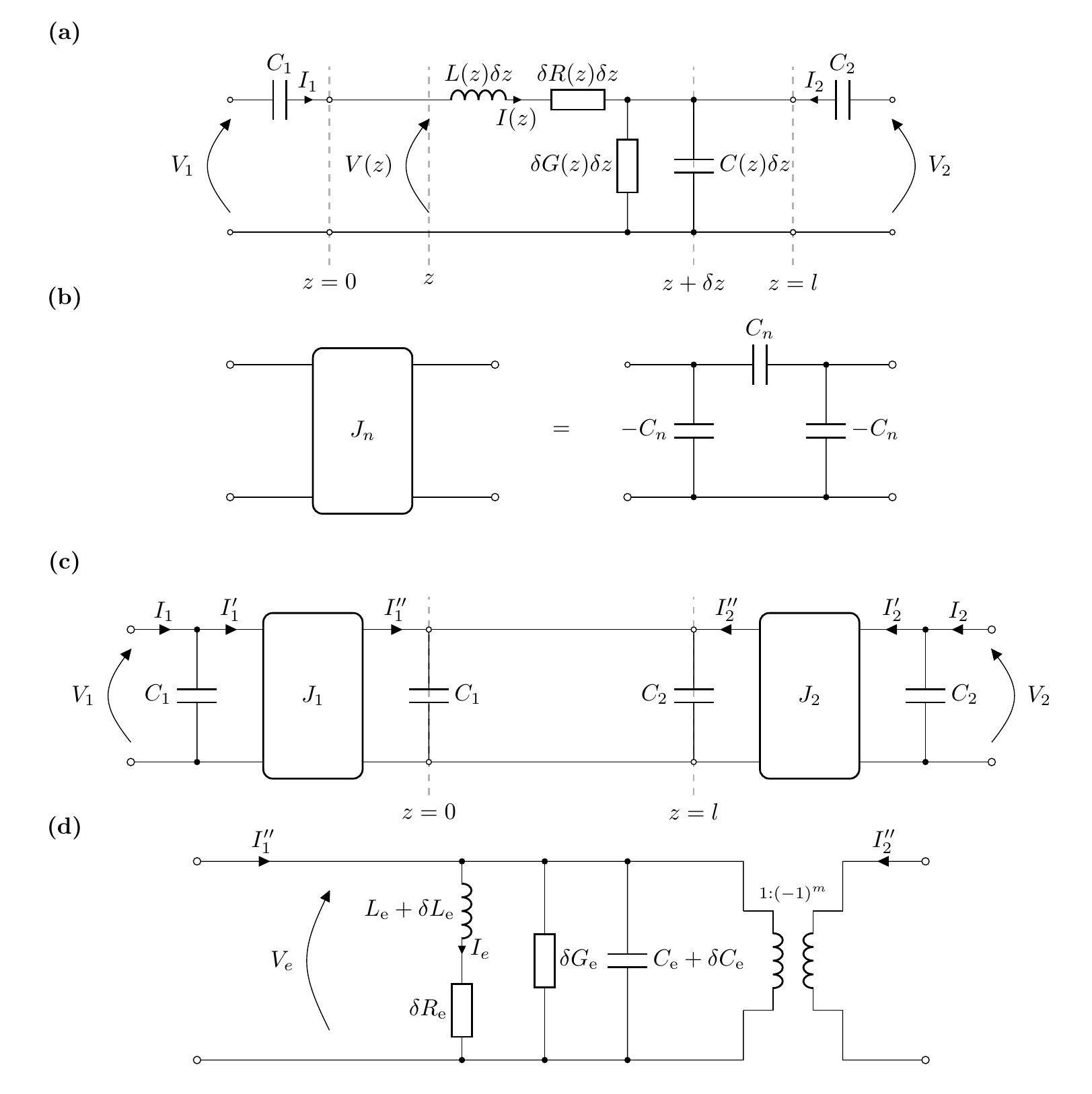}
\caption{\label{fig:hw_resonator}
A capacitively-coupled resonator transmission line resonator.
a) Basic circuit.
b) Admittance inverter.
c) Basic circuit redrawn to include admittance inverters.
d) Equivalent circuit of the transmission line for frequencies near the $m^\text{th}$-order resonance.
}
\end{figure*}

\subsubsection{Preliminaries}\label{sec:hw_analysis_preliminaries}

We will work throughout in terms of complex analytic signal representations $f(t)$ of the physical voltages and currents\,\cite{mandel1995optical}.
We will assume these have been normalised so that the physical signal is given by $2 \Re[f(t)]$.
The time-averaged product of two physical signals $\Re[f(t)]$ and $\Re[g(t)]$ is therefore given by $2 \Re[f^*(t) g(t)]$.

Consider a length $l$ of transmission line coupled to two microwave ports by series capacitors, as depicted in Figure \ref{fig:hw_resonator}(a).
At position $z$ along the line, as measured from $C_1$, the line has series inductance $L(z) = L_0 + \delta C(z)$ per unit length, series resistance $\delta R (z)$ per unit length, shunt capacitance $C(z) = C_0 + \delta C(z)$ per unit length and shunt conductance $\delta G (z)$ per unit length.
We will assume the line is only weakly lossy and non-uniform at the frequencies $v$ of interest, such that $|\delta L (z)| / L_0$, $|\delta C (z)| / C_0 \ll 1$, $|\delta R (z)|/(2 \pi \nu) L_0$ and $|\delta G|/(2 \pi \nu) C_0 \ll 1$.
$V(z, t)$ and $I(z, t)$ will be used to denote the voltage and current as a function of position and time along the line.

The input $a_n$ and output signals $b_n$ of the state-space representation at each port $n$ are related to the terminal plane voltage $V_n$ and current $I_n$ by
\begin{equation}\label{eqn:hw_def_a}
	a_n (t) = \frac{1}{\sqrt{2 Z_n}} \left\{ V_n (t) + Z_n I_n(t) \right\}
\end{equation}
and
\begin{equation}\label{eqn:hw_def_b}
	b_n (t) = \frac{1}{\sqrt{2 Z_n}} \left\{ V_n (t) - Z_n I_n(t) \right\}
\end{equation}
where $Z_n$ is the (real-valued) reference impedance at each port.
These are simply the time-domain forms of the Kurokawa power wave amplitudes that are more normally used in the frequency domain\,\cite{kurokawa1965power}.
With these definitions
\begin{equation}
	|a_n(t)|^2 - |b_n(t)|^2 = 2 \Re[I^*(t) V(t)] = P_n (t),
\end{equation}
as required.
$V_n$ and $I_n$ are given in terms of $a_n$ and $b_n$ by
\begin{equation}\label{eqn:hw_vn}
	V_n (t) = \left\{ a_n (t) + b_n (t) \right\} \sqrt{Z_n / 2}
\end{equation}
and
\begin{equation}\label{eqn:hw_in}
	I_n (t) = \left\{ a_n (t) - b_n (t) \right\} / \sqrt{2 Z_n}.
\end{equation}

The derivation of the state-space equations is significantly simplified by introducing a pair of fictitious admittance inverters\,\cite{pozar2011microwave} between the coupling capacitors and the resonator.
Figure \ref{fig:hw_resonator}(b) defines the composition of the admittance inverters, and Figure \ref{fig:hw_resonator}(c) shows the basic circuit redrawn to incorporate them.
It is straightforward to show that the new currents introduced must satisfy
\begin{equation}\label{eqn:hw_i1p}
	I'_1 = - C_1 \left( \frac{\partial V}{\partial t} \right)_{z=0},
\end{equation}
\begin{equation}\label{eqn:hw_i2p}
	I'_2 = - C_2 \left( \frac{\partial V}{\partial t} \right)_{z=l},
\end{equation}
\begin{equation}\label{eqn:hw_i1pp}
	I''_1 = C_1 \frac{d V_1}{d t},
\end{equation}
and
\begin{equation}\label{eqn:hw_i2pp}
	I''_2 = C_2 \frac{d V_2}{d t}.
\end{equation}
We will see that this step removes the need to consider separate dynamical equations for the voltages across the coupling capacitors.

\subsubsection{Equivalent circuit for the transmission line resonator}\label{sec:resonator_equivalent_circuit}

As an intermediate step in deriving the state equation, it is useful to develop an equivalent circuit representation of the transmission line section.
It is well known that a lightly capacitively-coupled transmission line circuit behaves like a parallel inductor-capacitor (LC) circuit for excitation frequencies near the line's half wave resonances\,\cite{pozar2011microwave}.
However, this analysis is normally carried out in the frequency domain and includes both the line and the coupling capacitors.
Here we will derive the equivalent picture in the time-domain by application of the Telegrapher's equations to the line section in Figure \ref{fig:hw_resonator}(c).

Based on the results of the previous section we want to determine $V(0)$ and $V(l)$ in terms of $I''_1$ and $I''_2$; (\ref{eqn:hw_i1pp}) and (\ref{eqn:hw_i2pp}) can be used to calculate $I''_n$ from port voltage $V_n$, then (\ref{eqn:hw_i1p}) and (\ref{eqn:hw_i2p}) used to find the corresponding port currents $I_1$ and $I_2$.
In general, a function $f(x)$ on an interval $0 < z < l$ can be represented as either a sine or cosine series:
\begin{align}
	f(z) &= \frac{c_0}{2} + \sum_{n=1}^\infty c_n \cos \Bigl( \frac{n \pi z}{l} \Bigr)
	= \sum_{n=1}^\infty b_n \sin \Bigl( \frac{n \pi z}{l} \Bigr) \label{eqn:sine_cosine_series} \\
	c_n &= \frac{2}{l} \int_0^l f(z) \cos \Bigl( \frac{n \pi z}{l} \Bigr) \, dz \label{eqn:cosine_coeffs} \\
	b_n &= \frac{2}{l} \int_0^l f(z) \sin \Bigl( \frac{n \pi z}{l} \Bigr) \, dz \label{eqn:sine_coeffs}.
\end{align}
Assume we are interested in exciting the $m^\text{th}$-order resonance at frequency $\nu_m$, where $m=1$ corresponds to the fundamental resonant mode.
In the case of a capacitively line with weak coupling ($2 \pi \nu_m C_n Z_n \ll 1$), the effective input impedance of the readout circuit as measured at the ends of the line will be very high, so we expect the end currents to be very small.
This would suggest it is most appropriate to expand $I(z)$ as a sine series and $V(z)$ as a cosine series (the latter for eventual consistency with the Telegrapher's equations).
Hence for excitation frequencies near $\nu_m$, we would expect the voltage and current on the line to have the approximate forms
\begin{equation}\label{eqn:hw_v_line}
	V(z, t) \approx V_e (t) \cos \left( \frac{m \pi z}{l} \right)
\end{equation}
and
\begin{equation}\label{eqn:hw_i_line}
	I(z, t) \approx \frac{2}{m \pi} I_e (t) \sin \left( \frac{m \pi z}{l} \right),
\end{equation}
based on a travelling wave expansion, where the factor of $2/m \pi$ has been included in (\ref{eqn:hw_i_line}) for later algebraic convenience.
We could equivalently argue (\ref{eqn:hw_v_line}) and (\ref{eqn:hw_i_line}) on the physical basis that we expect the line to behave like an open-ended resonator.
From (\ref{eqn:hw_v_line}) and (\ref{eqn:hw_i_line}) we see that the problem of determining $V(0, t)$ and $V(l, t)$ is equivalent to determining $V_e (t)$.

The equations describing the dynamical behaviour of the decomposition coefficients $V_e$ and
 $I_e$ follow from the Telegrapher's equations.
The latter are
\begin{equation}\label{eqn:te1}
	\{ C (z) + 2 C_1 \delta (z) + 2 C_2 \delta(z - l) \} \frac{\partial V}{\partial t} =
	- \frac{\partial I }{\partial z} - \delta G(z) V(z, t)
\end{equation}
and
\begin{equation}\label{eqn:te2}
	L (z) \frac{\partial I}{\partial t}
	= - \frac{\partial V }{\partial z} - \delta R(z) I(z, t)
\end{equation}
for the line as it appears in Figure \ref{fig:hw_resonator}(c).
A set of dynamical equations for the decomposition coefficients by substituting (\ref{eqn:hw_v_line}) and (\ref{eqn:hw_i_line}) into (\ref{eqn:te1}) and (\ref{eqn:te2}), then using (\ref{eqn:sine_coeffs}) and (\ref{eqn:cosine_coeffs}) to extract the decomposition coefficients.
The spatial derivative terms can be evaluated using the identities
\begin{equation}
\begin{aligned}
	&\int_0^l \left\{ \begin{matrix}
		\sin \left( \frac{m \pi z}{l} \right) \\
		\cos \left( \frac{m \pi z}{l} \right)
	\end{matrix} \right\} \frac{d f}{dz} \, dz \\
	&= \left\{ \begin{matrix}
		0 \\
		(-1)^m f(l) - f(0)
	\end{matrix} \right\}
	- \frac{m \pi}{l} \int_0^l \left\{ \begin{matrix}
		\cos \left( \frac{m \pi z}{l} \right) \\
		-\sin \left( \frac{m \pi z}{l} \right)
	\end{matrix} \right\} f(z) \, dz
\end{aligned}
\end{equation}
which follow from integration by parts.
Doing so, we obtain a pair of coupled dynamical equations:
\begin{align}
	&\{ L_e + \delta L_e \} \frac{d I_e}{dt}
	= +V_e (t) - \delta R_e I_e (t)
	\label{eqn:hw_ie_de} \\
	&\{ C_e + \delta C_e \} \frac{d V_e}{dt} \nonumber \\
	&=  -I_e (t) - \delta G_e V_e (t) + I''_1 (t) + (-1)^m I''_2 (t)
	\label{eqn:hw_ve_de},
\end{align}
where $C_e = C_1 + C_2 + C_0 l / 2$, $L_e = (2 / m \pi)^2 L_0 l / 2$,
\begin{equation}\label{eqn:def_ce_and_ge}
	\left\{ \begin{matrix}
		\delta C_e \\
		\delta G_e
	\end{matrix} \right\}
	= \int_0^l \left\{ \begin{matrix}
		\delta C (z) \\
		\delta G (z)
	\end{matrix} \right\}
	\cos^2 \left( \frac{m \pi z}{l} \right) \, dz
\end{equation}
and
\begin{equation}\label{eqn:def_le_and_re}
	\left\{ \begin{matrix}
		\delta L_e \\
		\delta R_e
	\end{matrix} \right\}
	= \left( \frac{2}{m \pi} \right)^2 \int_0^l \left\{ \begin{matrix}
		\delta L (z) \\
		\delta R (z)
	\end{matrix} \right\}
	\sin^2 \left( \frac{m \pi z}{l} \right) \, dz.
\end{equation}
(\ref{eqn:hw_ve_de}) and (\ref{eqn:hw_ie_de}) can be recognised as the dynamical equations describing the voltages and current labeled $V_e$ and $I_e$ in the parallel LC circuit shown in Figure \ref{fig:hw_resonator}(d).
$L_e + \delta L$, $C_e + \delta C_e$, $\delta R_e$ and $\delta G_e$ can be seen to be the equivalent inductance, capacitance, resistance and conductance in this representation.
The output transformer is necessary to account for the factor of $(-1)^n$ in (\ref{eqn:hw_ve_de}).
In the analysis that follows it will be sufficient to model the transmission line by the parallel LC circuit, Figure \ref{fig:hw_resonator}(d), as described by the coupled pair of equations (\ref{eqn:hw_ie_de}) and (\ref{eqn:hw_ve_de}).
Hence we have derived the normal parallel LC model for a half-wave resonator line interacting with other circuits.

Before moving on it is useful to briefly comment on some of the less intuitive aspects of the model described.
Firstly, it may at first seem physically counter-intuitive that the resonant mode described by (\ref{eqn:hw_ie_de}) and (\ref{eqn:hw_ve_de}), which results in no currents at the end of the line, is in fact driven by end currents.
Secondly, the question then arises that if we excite this mode and it has no end currents, how can the original boundary condition imposed by the end currents be met?
In the case of the first problem, mathematically the decomposition must necessarily leave the end currents undefined; they are constraints and the decomposition describes degrees of freedom.
The second problem is resolved by considering the full sine series representation, i.e. (\ref{eqn:sine_cosine_series}).
A sine series can be made a finite value an arbitrarily small (but non-zero) distance from the ends of the domain by adding sufficient terms, ensuring continuity with the boundary conditions.
If the full voltage and current decompositions are substituted into the Telegrapher's equations, it is found that any degree of coupling results in the excitation of all of the spatial harmonics to some extent, if even if the drive frequency is far from the natural resonant frequency.
Physically, the coupling introduces loss in each mode that broadens its spectral response.
The resultant pattern of harmonic excitation is such that it ensures the continuity with the impressed currents.

\subsubsection{Prototype state equation}\label{sec:hw_prototype_state_equation}

To derive the state-equation in the state-space representation it is necessary to define a set of complex mode amplitudes.
The average total energy stored on the line is given by
\begin{equation}
\begin{aligned}
	U (t) &= \int_0^l C(z) | V(z, t) |^2 + L (z) | I (z, t) |^2 \, dz \\
	&\approx C_e |V_e (t)|^2 + L_e |I_e (t)|^2 .
\end{aligned}
\end{equation}
Hence the correct amplitudes are some linear combination of $\sqrt{C_e} V_e$ and $\sqrt{L_e} I_e$ which decouples (\ref{eqn:hw_ve_de}) and (\ref{eqn:hw_ie_de}).

Consider mode amplitudes $u_+$ and $u_-$ as defined by
\begin{equation}\label{eqn:def_up}
	u_+ (t) = \frac{1}{\sqrt{2}} \Bigl[
		\sqrt{C_e} V_e (t) + i \sqrt{L_e} I_e (t)
	\Bigr].
\end{equation}
and
\begin{equation}\label{eqn:def_un}
	u_- (t) = \frac{1}{\sqrt{2}} \Bigl[
		\sqrt{C_e} V_e (t) - i \sqrt{L_e} I_e (t)
	\Bigr].
\end{equation}
It is straightforward to show these satisfy the requirement $U(t) = |u_+ (t)|^2 + |u_- (t) |^2 = U(t)$.
(\ref{eqn:hw_ve_de}) and (\ref{eqn:hw_ie_de}) can be rewritten in terms of (\ref{eqn:def_up}) and (\ref{eqn:def_un}) as
\begin{equation}\label{eqn:prototype_state_equation}
\begin{aligned}
	\frac{d \mathbf{u}}{dt}
	&= 2 \pi i \nu_m
	\left( \begin{smallmatrix}
		1 + \delta \chi_1 & \chi_2 \\
		-\chi_2^* & - (1 + \delta \chi^*_1)
	\end{smallmatrix} \right) \cdot \mathbf{u} \; + \\
	& \frac{1}{4 \pi i \nu_m}
	\left( \begin{smallmatrix}
		k_1 & k_2 \\
		k_1 & k_2
	\end{smallmatrix} \right) \cdot
	\left\{ \frac{d \mathbf{a}}{dt}
	+ \frac{d \mathbf{b}}{dt}  \right\}
\end{aligned}
\end{equation}
where $\mathbf{u} = (u_+, u_-)$, $\mathbf{a} = (a_1, a_2)$ and $\mathbf{b} = (b_1, b_2)$.
(\ref{eqn:hw_i1pp}), (\ref{eqn:hw_i2pp}) and (\ref{eqn:hw_vn}) have been used to rewrite the $I''_n$ in terms of $a_n$ and $b_n$.
The definitions of the other quantities in (\ref{eqn:prototype_state_equation}) are
\begin{equation}\label{eqn:def_vn}
	\nu_m = \frac{1}{2 \pi \sqrt{L_e C_e}}
	= \frac{m}{2 l \sqrt{L_0 C_0}}
	= \frac{m c_l}{2 l},
\end{equation}
\begin{equation}\label{eqn:def_ze}
	Z_e = \sqrt{\frac{L_e}{C_e}}
	= \frac{2 \eta}{m \pi},
\end{equation}
\begin{equation}\label{eqn:def_chi1}
	\delta \chi_1 = \frac{1}{2} \left\{
		-\frac{\delta L_e}{L_e} - \frac{\delta C_e}{C_e}
		+ \frac{i \delta R_e}{Z_e} + i Z_e \delta G_e
	\right\},
\end{equation}
\begin{equation}\label{eqn:def_chi2}
	\delta \chi_2 = \frac{1}{2} \left\{
		-\frac{\delta L_e}{L_e} + \frac{\delta C_e}{C_e}
		- \frac{i \delta R_e}{Z_e} + i Z_e \delta G_e
	\right\},
\end{equation}
\begin{equation}
	k_1 = i \sqrt{\frac{(2 \pi \nu_m C_1)^2 Z_1}{C_e}}
\end{equation}
and
\begin{equation}
	k_2 = (-1)^m i \sqrt{\frac{(2 \pi \nu_m C_2)^2 Z_2}{C_e}} .
\end{equation}
$c_l$ and $\eta$ are the wave speed and characteristic impedance of the transmission line.
(\ref{eqn:prototype_state_equation}) shows that the choice of mode amplitudes decouples the equations to first order in $\delta \chi_2$.
$u_+$ is the amplitude of the mode with resonant frequency $\nu_m$, while $\nu_m$ is the same for the mode with resonant frequency $-\nu_m$.

\subsubsection{Prototype output equation}\label{sec:hw_prototype_output_equation}

The prototype output equation follows by considering conservation of current at the nodes labelled $A$ and $B$ in Figure \ref{fig:hw_resonator}(c).
This yields a dynamical equation
\begin{equation}
	I_n = C_n \frac{d V_n}{dt}
	+ I'_n (t)
	= C_n \frac{d V_n}{dt}
	- C_n \begin{cases}
		(\partial V / \partial t)_{z=0} & n = 1 \\
		(\partial V / \partial t)_{z=l} & n = 2
	\end{cases}
\end{equation}
for each port.
Rewriting $V_n$, $I_n$ and $V$ in terms of $a_n$, $b_n$, $u_+$ and $u_-$ and rearranging, we obtain
\begin{equation}
\begin{aligned}
	& \mathbf{b}
	+ \left( \begin{smallmatrix}
		C_1 Z_1 & 0 \\
		0 & C_2 Z_2
	\end{smallmatrix} \right) \cdot
	\frac{d \mathbf{b}}{dt} =\\
	&\mathbf{a}
	- \left( \begin{smallmatrix}
		C_1 Z_1 & 0 \\
		0 & C_2 Z_2
	\end{smallmatrix} \right) \cdot
	\frac{d \mathbf{a}}{dt}
	+ \\
	&\frac{1}{2 \pi i \nu_m}
	\mathbf{k}
	\frac{du_+}{dt}
	+ \frac{1}{2 \pi i \nu_m}
	\mathbf{k}
	\frac{du_-}{dt}
\end{aligned}
\end{equation}
for $\mathbf{k} = (k_1, k_2)$.

Finally, we will make the assumption of weak coupling: $|2 \pi \nu_m C_n Z_n| \ll 1$.
If this is not the case, the coupling capacitors short out the resonator response at $\nu_n$ and the behaviour is uninteresting.
With this approximation we obtain
\begin{equation}\label{eqn:prototype_output_equation}
	\mathbf{b}
	\approx \mathbf{a}
	+ \frac{1}{2 \pi i \nu_m}
	\mathbf{k}
	\left\{ \frac{du_+}{dt}
	+ \frac{du_-}{dt}
	\right\}.
\end{equation}

\subsubsection{Final state-space representation}\label{sec:hw_state_space_representation}

To convert (\ref{eqn:prototype_state_equation}) and (\ref{eqn:prototype_output_equation}) to the state-space representation in Section \ref{sec:resonator_model} two additional assumptions need to be made.
Firstly, assume the spectrum of the input signals is narrow band and centred around $\nu_m$.
The output signals will then have similar spectral content and we may approximate
\begin{equation}\label{eqn:narrow_band_approx}
	\frac{df}{dt} \approx 2 \pi \nu_m f(t)
\end{equation}
for all signals.
Secondly, assume the Q-factor is sufficiently high that $u_-$ is not excited significantly when the input signal has positive frequency, i.e. that we may assume $u_-$ zero.
This is equivalent to the single-pole approximation usually made to the frequency-domain response of a resonator.

With the approximations allowed by these assumptions (\ref{eqn:prototype_output_equation}) immediately becomes (\ref{eqn:coupled_output_equation}), with
\begin{equation}
	S = \left( \begin{smallmatrix}
		0 & 1 \\
		1 & 0
	\end{smallmatrix} \right).
\end{equation}
(\ref{eqn:prototype_state_equation}) becomes
\begin{equation}
	\frac{du_+}{dt}
	= 2 \pi \nu_m \left\{ 1 + \frac{d \fres}{\fres} + \frac{i}{2 Q_i} \right\} u_+
	+ \frac{1}{2} \mathbf{k}^T \cdot \left\{ \mathbf{a} + \mathbf{b} \right\},
\end{equation}
where we may now identify
\begin{equation}\label{eqn:hw_fres_shift}
\begin{aligned}
	&\frac{d \fres}{\fres}
	= \frac{1}{2} \left\{ \frac{\delta L_e}{L_e} + \frac{\delta C_e}{C_e} \right\} \\
	&= \frac{1}{l} \int_0^l
		\frac{\delta L(z)}{L_0} \cos^2 \left( \frac{m \pi z}{l} \right)
		+ \frac{\delta C(z)}{C_0} \sin^2 \left( \frac{m \pi z}{l} \right)
	\, dz
\end{aligned}
\end{equation}
and
\begin{equation}\label{eqn:hw_qi}
\begin{aligned}
	&\frac{1}{Q_i}
	= \frac{\delta R_e}{Z_e} + Z_e \delta G_e \\
	&= \frac{2}{m \pi} \int_0^l
		\frac{\delta R (z)}{\eta} \cos^2 \left( \frac{m \pi z}{l} \right)
		+ \eta \delta G (z) \sin^2 \left( \frac{m \pi z}{l} \right)
	\, dz.
\end{aligned}
\end{equation}
(\ref{eqn:coupled_state_equation}) then follows by using (\ref{eqn:coupled_output_equation}) to substitute for $\mathbf{b}$.
For completeness we note
\begin{equation}\label{eqn:hw_qc}
\begin{aligned}
	\frac{1}{Q_c}
	&= \frac{(2 \pi \nu_m) \{ C_1^2 Z_1 + C_2^2 Z_2 \}}{C_e}.
\end{aligned}
\end{equation}

\subsubsection{Device-averaged quantities}\label{sec:dev_averaged_quantities}

To illustrate the concept of device-averaged quantities, consider the case where the series resistance per unit length is proportional to local quasiparticle density: $\delta R (z) = R_* \nqp (z) / n_*$.
It follows from (\ref{eqn:hw_qi}) that we may write
\begin{equation}\label{eqn:def_qi}
	\frac{1}{Q_i} = \frac{\nqp' l R_* }{m \pi n_* \eta},
\end{equation}
where
\begin{equation}\label{eqn:dev_averaged_qp}
	\nqp' = \frac{2}{l} \int_0^l \nqp(z) \cos^2 \left( \frac{m \pi z}{l} \right) \, dz
\end{equation}
is what is meant by the device-averaged quasiparticle density.
Note that the presence of the weighting function in the integral, which means $Q_i$ more sensitive to quasiparticles generated in the centre of the resonator then at the ends.

\section{Illustrative model}\label{sec:illustrative_model_sm}

\subsection{Stationary points of $|1 - \Gamma_p|^2$ with respect to $y_p$}\label{sec:stationary_points}

We begin by finding the derivative of $|1 - \Gamma_p|^2$ with respect to $y_p$.
Using (\ref{eqn:gamma_p_norm}), it follows that
\begin{equation}\label{eqn:sp_deriv_intermediate}
	\frac{d|1 - \Gamma_p|^2}{dy_p}
	= \frac{d}{d y_p} \biggl\{ \frac{4 q_r^2}{|p|^2} \biggr\}
	= - \frac{8 q_r^2}{|p|^4} \Re \biggl[ p^* \frac{d p}{d y_p} \biggr]
\end{equation}
in general.
To calculate the derivative of $p$, we use the fact that
\begin{equation}\label{eqn:sp_p}
	p = 1 + 2 i y_p \mp \frac{2 i a_* e^{\pm i \phi}}{|p|^2},
\end{equation}
in the case of the illustrative model, which is obtained by using (\ref{eqn:U0_norm}) to substitute for $|u|^2 / U_*$ in (\ref{eqn:ill_mod_p}).
Differentiating (\ref{eqn:sp_p}) with respect to $y_p$  gives
\begin{equation}\label{eqn:deriv_p_wrt_yp}
	\frac{d p}{d y_p}
	= 2 i \left\{ 1 \mp \frac{a_* e^{\pm i \phi}}{4 q_r^2}
	\frac{d| 1 - \Gamma_p|^2}{d y_p} \right\},
\end{equation}
where we have used (\ref{eqn:sp_deriv_intermediate}) to rewrite the derivative of $1 / |p|^2$ in terms of the derivative of $|1 - \Gamma_p|^2$.
(\ref{eqn:sp_deriv_intermediate}) and (\ref{eqn:deriv_p_wrt_yp}) can then be solved together to yield
\begin{equation}\label{eqn:sp_deriv}
	\frac{d |1 - \Gamma_p|^2}{d y_p}
%	= -\frac{16 q_r^2 \Im[p] }{|p|^4}
%	\left[ 1 \mp \frac{4 q_r \Im[p e^{\mp \i \phi}]}{|p|^2}\right]^{-1}
	= -\frac{16 q_r^2 \Im[p] }{|p|^4 \mp 4 q_r \Im[p e^{\mp i \phi}]} .
\end{equation}

(\ref{eqn:sp_deriv}) is zero at the stationary points of $|1 - \Gamma_p|^2$ with respect to $y_p$.
It is straightforward to see this requires either (a) $\Im[p] = 0$ or (b) $|p| \rightarrow \infty$.
Condition (b) can only be achieved either at a limit, e.g. $|y_p| \rightarrow \infty$, or in the case of extreme nonlinear behaviour.
Condition (a) is instead more likely for an isolated stationary point in a sweep.
If indeed $\Im[p] = 0$ at the stationary point, then it follows by taking the imaginary part of (\ref{eqn:sp_p}) that it must be that
\begin{equation}
	y_p = \pm \frac{a_* \cos \phi}{|p|^2},
\end{equation}
which is the result used in the main text.
It is also worth noting that $\Im[p] = 0$ implies
\begin{equation}
	y_p = \Re[g(v)] = \frac{Q_r \delta f_0}{f_0}
\end{equation}
or, equivalently,
\begin{equation}
	f_p = f_0 + \delta f_0,
\end{equation}
so at the stationary point the pump frequency is equal to the shifted resonant frequency.

\subsection{Signal gain in the vicinity of a switching point}\label{sec:gain_near_sp}

Let $\kps$, $\as$ and $\ks$ denote the values of $k_p$, $a$ and $k_s$, respectively, at the switching point of interest.
We will investigate how the signal gain $G_s$, as given by (\ref{eqn:general_signal_power_gain}), behaves as a function of $\delta k_p = k_p - \kps$ and $\delta a = a - a_s$ in three different cases:
\begin{mycases}
\item The switching point is the critical point, so $\kps = \sqrt{3} / 2$, $\ks = 1 / (2 \sqrt{3})$ and $\as = 4 / (3 \sqrt{3})$.
This corresponds to the region below the apex of the dotted and dashed (red) lines in the righthand panels of Figures \ref{fig:gain_contour_plot_phi_zero} and \ref{fig:gain_contour_plot_phi_10deg}.
\item The switching point is from the low- to high-energy ($y_1$ in Figure \ref{fig:generic_switching_curve}) and $k_p \gg \sqrt{3}/2$, so we may approximate (\ref{eqn:sp_k_vals}) and (\ref{eqn:a_from_k_and_k_p}) as $\ks = 2 \kps / 3$ and $\as = 16 \kps^3 / 27$.
This corresponds to the region immediately to the left of the dashed (red) lines in the right-hand panels of Figures \ref{fig:gain_contour_plot_phi_zero} and \ref{fig:gain_contour_plot_phi_10deg} at large $|y_p|$.
\item The switching point is from the high- to low-energy ($y_2$ in Figure \ref{fig:generic_switching_curve}) and $k_p \gg \sqrt{3}/2$, so we may approximate (\ref{eqn:sp_k_vals}) and (\ref{eqn:a_from_k_and_k_p}) as $\ks = 0$ and $\as = \kps$.
This corresponds to the region immediately to the right of the dotted (red) lines in the left-hand panels of Figures \ref{fig:gain_contour_plot_phi_zero} and \ref{fig:gain_contour_plot_phi_10deg} at large $|y_p|$.
\end{mycases}

Using (\ref{eqn:sm_p}) and (\ref{eqn:sm_q}), we can express the signal gain as a function of $k$ and $k_p$:
\begin{equation}
	G_s = \frac{4 Q_r^2}{Q_c^2}
	\frac{1 + 16 k^2 - 16 k_p k - 4 k_p^2}{(12 k^2 - 8 k_p k +1)^2}.
\end{equation}
The gain becomes singular at the switching point because the denominator of this expression goes to zero.
To understand the behaviour near this point, we would ideally like to Taylor expand the denominator to at least first order in the independent variables $\delta k_p$ and $\delta a$.
However, it is not possible to do so because the necessary derivatives
\begin{equation}
	\frac{\partial k}{\partial k_p} = \frac{4 k^2 + 1}{12 k^2 - 8 k_p k + 1}
\end{equation}
and
\begin{equation}
	\frac{\partial k}{\partial a} = -\frac{1}{12 k^2 - 8 k_p k + 1}
\end{equation}
are themselves also singular at the switching point.
Instead we can expand the denominator in terms of $\delta k = k - \ks$ and $\delta k_p$ and approximate the numerator by its value at the singular point to give
\begin{equation}\label{eqn:gs_approx}
	G_s \approx \frac{\qr^2}{4} \frac{(\ks - \kps)^2}{\{ 3 \delta k^2 / 2 + (3 \ks - \kps) \delta k
	- \ks \delta k_p - \delta k_p \delta k_s \}^2}
\end{equation}
The problem then becomes one of determining an expression for $\delta k$ in terms of the independent variables.

We can find such an expression by considering the operating point equation, (\ref{eqn:sm_state_equation}).
The latter can be rewritten as $f(k_p, a, k) = 0$,
where
\begin{equation}
	f(k_p, a, k) = k - k_p + \frac{a}{1 + 4 k^2}.
\end{equation}
We can expand $f(k_p, a, k)$ as a multivariate Taylor series in $\delta k_p$, $\delta a$ and $\delta k$ about a switching point as:
\begin{equation}\label{eqn:f_taylor_expansion}
\begin{aligned}
	&f(\kps + \delta k_p, \as + \delta a, \ks + \delta k) \\
	&= - \delta k_p + \frac{\kps - \ks}{\as} \delta a \\
	&- \frac{1}{\as} \delta k \delta a + \frac{4 (12 \ks^2 - 1) \as}{(1 + 4 \ks^2)^3} \delta k^2 \\
	&+ \frac{64  (1 - 4 \ks^2) \ks \as}{(1 + 4 \ks^2)^4} \delta k^3
	+ \frac{8 (12 \ks^2 - 1)}{3 (1 + 4 \ks^2)^3} \delta k^2 \delta a  + \dots.
\end{aligned}
\end{equation}
At any valid operating near the switching point $f(\kps + \delta k_p, \as + \delta a, \ks + \delta k)$ must be zero.
Hence by setting (\ref{eqn:f_taylor_expansion}) to zero we can obtain an equation relating $\delta k_p$, $\delta k$ and $\delta a$ up to different orders depending on the circumstances.
Note that in general $\delta k$ will depend on a root of $\delta k_p$ or $\delta a$, which explains why the Taylor series expansion of the denominator did not succeed.

As an example, consider the simplified case where $\delta a = 0$.
Setting (\ref{eqn:f_taylor_expansion}) equal to zero, we then obtain
\begin{equation}\label{eqn:da_zero}
	0 \approx -\delta k_p +\begin{cases}
		3 {\delta k}^3 & \text{Case 1} \\
		9 \delta k^2 / (4 \kps) & \text{Case 2} \\
		-4 \kps \delta k^2 & \text{Case 3}
	\end{cases}
\end{equation}
to leading order in $\delta k$, where we have also rewritten $\as$ in terms of $\kps$.
(\ref{eqn:da_zero}) rearranges to
\begin{equation}\label{eqn:dk_da_zero}
	\delta k = \begin{cases}
		\sqrt[3]{\frac{\delta k_p}{3}} & \text{Case 1} \\
		\frac{2}{3} \sqrt{\kps \delta k_p} & \text{Case 2} \; (\text{and } \delta k_p \geq 0) \\
		\tfrac{1}{2} \sqrt{-\frac{\delta k_p}{\kps}} & \text{Case 3} \; (\text{and } \delta k_p \leq 0).
	\end{cases}
\end{equation}
In each case $\delta k$ is proportional to a root of $\delta k_p$, hence we would expect $\delta k \gg  {\delta k}^2 \gg \delta k_p$ for $\delta k_p \ll 1$.
Therefore, we can approximate the signal gain as
\begin{equation}
	G_s \approx \qr^2 \begin{cases}
		\frac{\sqrt[3]{3}}{ 9\delta k_p^{4/3}} & \text{Case 1} \\
		\frac{1}{16 \kps |\delta k_p|} & \text{Case 2} \; (\text{and } \delta k_p \geq 0) \\
		\frac{\kps}{|\delta k_p|} & \text{Case 3} \; (\text{and } \delta k_p \leq 0).
	\end{cases}
\end{equation}
In the case of Figure \ref{fig:gain_contour_plot_phi_zero} we have $k_p = -y_p$, so
\begin{equation}
	G_s \approx \qr^2 \begin{cases}
		\frac{\sqrt[3]{3}}{ 9\delta y_p^{4/3}} & \text{Case 1} \\
		\frac{1}{16 y_1 |\delta k_p|} & \text{Case 2} \; (\text{and } \delta y_p \leq 0) \\
		\frac{y_2 }{|\delta y_p|} & \text{Case 3} \; (\text{and } \delta y_p \geq 0),
	\end{cases}
\end{equation}
as used in the main text.

Although it is not presented in the main text, we can also obtain a general analytic approximation for the gain near the critical point (Case 1) outside the bifurcation region.
Setting (\ref{eqn:f_taylor_expansion}) to zero and truncating the series at the third-order yields the following cubic equation for $\delta k$ in Case 1:
\begin{equation}\label{eqn:dk_cubic}
	{\delta k}^3 - \frac{\sqrt{3}}{4} \delta a \delta k + \frac{1}{4} \delta a - \frac{1}{3} \delta k_p = 0.
\end{equation}
This equation can have either one or three physical, real-valued, solutions for $\delta k$.
Intuitively, we would expect the latter situation to correspond to that where the operating point lies in the bifurcation regime and so the device can approach the critical point from one of three states, each with different behaviour (two stable and one unstable).
The former case, with a single solution, should correspond to approaching the critical point from outside the bifurcation regime, i.e. region (i) in Figure \ref{fig:gain_contour_plot_phi_zero}.
Under the assumption of single solution for the cubic equation, we can use Cardano's formula on (\ref{eqn:dk_cubic}) to give
\begin{equation}\label{eqn:dk_general_case_1}
	\delta k = \sqrt[3]{-\frac{s}{2} + \sqrt{\frac{s^2}{4} + \frac{r^3}{27}}}
	+ \sqrt[3]{-\frac{s}{2} - \sqrt{\frac{s^2}{4} + \frac{r^3}{27}}}
\end{equation}
where
\begin{equation}
	r = -\frac{\sqrt{3}}{4} \delta a
\end{equation}
and
\begin{equation}
	s =  \frac{\delta a}{4} - \frac{\delta k_p}{3}.
\end{equation}
(\ref{eqn:dk_general_case_1}) can be used in (\ref{eqn:gs_approx}) to calculate the gain in the general case.

\end{document}